\documentclass[%
 reprint,
 superscriptaddress,
 preprintnumbers,
 nofootinbib,
 amsmath,amssymb,
 aps,
 prl,
]{revtex4-1}

\usepackage{algorithm}
\usepackage{algpseudocode}
\usepackage{psfrag,slashed,cancel,array,graphicx,todonotes}
\usepackage{subfigure}
\usepackage[labelfont=bf,labelsep=quad,justification=justified]{caption}
\usepackage[utf8]{inputenc}
\usepackage{mathtools}
\usepackage{mathrsfs}
\usepackage{multirow}
\usepackage{braket}
\usepackage{slashed}
\usepackage{booktabs} %nice tables
\usepackage[normalem]{ulem}
\usepackage{slashed}
\usepackage{hyperref}

\hypersetup{pdftitle={Analytical Soft Functions for Heavy-Quark Final States at Hadron Colliders},pdfcreator={},linkcolor=[rgb]{0.15,0.35,0.75},colorlinks=true,citecolor=[rgb]{0.675,0,0.2},urlcolor=[rgb]{0.15,0.35,0.65}}

\newcommand{\nn}{\nonumber}

\newcommand{\cT}{{\mathcal T}}

\newcommand{\gcusp}{\gamma_{\mathrm{cusp}}}

\newcommand{\bmT}{{\mathbf T}}

\begin{document}

\title{Analytical Soft Functions for Heavy-Quark Final States at Hadron Colliders}

\preprint{CERN-TH-2025-244}

\author{Ze Long Liu$^a$}\email{liuzelong@ihep.ac.cn}
\author{Pier Francesco Monni$^b$}\email{pier.monni@cern.ch}

\affiliation{${}^a$ Theoretical Physics Division, Institute of High Energy Physics, Chinese Academy of Sciences, Beijing 100049, China}

\affiliation{${}^b$ CERN, Theoretical Physics Department, CH-1211 Geneva 23, Switzerland}

\begin{abstract}
We present the first computation of the complete two-loop, fully-differential soft function describing the production of a heavy–quark pair in association with a color-singlet system at hadron colliders. This result constitutes one of the most complex soft functions known to date and it is obtained in closed analytic form for generic multi-dimensional kinematics. This allows us to obtain novel analytic results for the transverse-momentum-dependent and threshold soft functions 
in this class of processes. We further obtain a decomposition of the soft function into dipole and tripole color correlators, thereby supplying essential building blocks for processes involving a heavy–quark pair produced together with additional light jets at both hadron and lepton colliders. These results represent a key ingredient for advancing precision predictions for heavy-quark physics at the LHC.
\end{abstract}

\maketitle

%================
%=== Introduction ===
%================
 \paragraph*{Introduction.---}  
The production of heavy-quark pairs ($Q\bar Q$) in Quantum Chromodynamics (QCD), with or without an accompanying color-singlet state $V$, is a cornerstone of the physics program at high-energy colliders. Both at the Large Hadron Collider (LHC) and at future lepton facilities, these processes provide essential insight into the dynamics of top and bottom quarks, as well as access to their couplings to the Higgs and electroweak gauge bosons (cf.~e.g.~\cite{Janot:2015yza,Vos:2016til,CMS:2021klw,CMS:2022tkv,ATLAS:2023eld,CMS:2024fdo,ATLAS:2024gth,CMS:2024mke,ATLAS:2024moy,FCC:2025lpp}).

Accurate theoretical predictions are challenged by large infrared logarithms that spoil the convergence of fixed-order QCD calculations and require all-order resummation. For many collider observables, resummation is organized through factorization theorems, which capture the leading-power behavior of the observable in infrared regimes. Prominent examples at lepton colliders include the threshold factorization for $Q\bar Q(V)$ production~\cite{vonManteuffel:2014mva,Gao:2014nva,Gao:2014eea}. At hadron colliders, analogous factorizations govern the transverse-momentum spectrum of the $Q\bar Q(V)$ system~\cite{Zhu:2012ts,Li:2013mia,Catani:2014qha,Catani:2018mei,Ju:2024xhd}, or the cross section for $Q\bar Q(V)$ production in the threshold limit where the accompanying radiation is soft~\cite{Ahrens:2010zv}. These theorems also underpin slicing approaches to higher-order calculations~\cite{Gao:2014nva,Gao:2014eea,Catani:2019iny,Catani:2022mfv,Buonocore:2022pqq,Devoto:2024nhl,Buonocore:2025fqs} and their matching to parton showers~\cite{Mazzitelli:2020jio,Mazzitelli:2021mmm,Mazzitelli:2024ura}.

A key ingredient in all such factorization theorems is the soft function, which encodes the effect of soft emissions from the eikonalized trajectories of energetic QCD particles (either light partons or heavy quarks), represented by Wilson lines. Notably, the soft functions relevant for threshold and transverse-momentum-dependent (TMD) observables can be obtained from a more general soft function differential in the total momentum of soft radiation~\cite{Li:2016axz,Li:2016ctv}. This \textit{fully differential} soft function for two light-like Wilson lines has been known for some time through three loops~\cite{Li:2016ctv}, whereas its massive counterpart, with two time-like Wilson lines, has been recently obtained at the two-loop level~\cite{Liu:2024hfa}.
While the latter calculation is necessary for the accurate description of the production of a $Q\bar Q(V)$ final state at lepton colliders, the hadron-collider counterpart is considerably more complex due to the presence of initial-state radiation. In the context of the soft function, this is represented by two additional light-like Wilson lines that describe the energetic incoming partons.
In this case, the two-loop computation of the threshold soft function for $Q\bar{Q}$ production has been carried out in Refs.~\cite{Czakon:2013hxa,Wang:2018vgu}, though the analogous quantity for $Q\bar{Q}V$ is currently unknown. In the TMD case, numerical calculations have been performed for the corresponding soft function averaged over the azimuthal angle of the final-state system in Refs.~\cite{Angeles-Martinez:2018mqh,Catani:2023tby,Devoto:2025eyc}, and an analytic result has been derived for $Q\bar{Q}$ produced at threshold~\cite{Shao:2025dzw}.

In this Letter we present the first analytic calculation of the complete two-loop, fully-differential soft function for $Q\bar{Q}V$ production at hadron colliders. The information encoded in this object is essential for theoretical investigations into the structure of multi-leg soft functions with both light-like and time-like Wilson lines. From a phenomenological viewpoint, it is relevant for precision phenomenology involving heavy-quark final states, for instance, in the context of higher-order calculations of azimuthal asymmetries~\cite{Catani:2017tuc} in final states involving pairs of heavy quarks.
The result~\cite{zenodo} is expressed in terms of Goncharov polylogarithms (GPLs) as a function of the generic kinematics of both the $Q\bar{Q}V$ final state and of the additional two light-like directions which correspond to the incoming legs in hadron collisions. This makes it one of the most complex soft functions known analytically. We will also discuss and present its decomposition into dipole and tripole color-correlator structures, which directly provide the most complicated contributions to more general soft functions for processes where the $Q\bar{Q}V$ system is accompanied by light QCD jets in the final state, whose properties are being explored at the LHC.

%=====================================
%=== General form of anomalous dimensions ===
%=====================================
\paragraph*{Soft functions in factorization theorems.---} 
We study the production of a heavy-quark pair in association with a color-singlet system in hadronic collisions,
\[
N_1(P_1)+N_2(P_2)\to Q(p_3)+\bar Q(p_4)+V(p_V)+X.
\]
Factorization theorems offer a powerful framework for resumming the large logarithmic enhancements that arise in QCD to all orders in the strong coupling $\alpha_s$. A prominent example is the factorization of the transverse--momentum distribution $p_\perp \equiv |\vec p_\perp|$ of the $Q\bar Q V$ system with respect to the beam direction in the regime $p_\perp \ll M$, where $M =\sqrt{ (p_3+p_4+p_V)^2}$ denotes the invariant mass of the final state and is of order the partonic center-of-mass energy $\sqrt{\hat s}$~\cite{Collins:1984kg,Ji:2004wu,Becher:2010tm,Chiu:2011qc,Zhu:2012ts,Li:2013mia,Catani:2014qha}. Within soft-collinear effective theory (SCET)~\cite{Bauer:2001yt,Bauer:2002nz,Beneke:2002ph}, the leading-power factorization formula in the small-$p_\perp$ limit, characterized by the expansion parameter $\lambda = |\vec p_\perp|/M$, takes the form
\begin{align}\label{eq:TMDfac}
\frac{d\sigma}{d^2 \vec{p}_\perp d\Omega} = &\int \!\! dz_1 \int \!\!d z_2 
\int \!\! d^2 \vec{x}_{\perp} \, e^{i \vec{p}_\perp\cdot \vec{x}_\perp}\\
&\times\sum_{i,j} B_{i/N_1}(z_1,\vec{x}_\perp,\mu,\nu) B_{j/N_2}(z_2,\vec{x}_\perp,\mu,\nu) \nn\\
&\times{\rm tr} \langle {\boldsymbol {\mathcal H}}_{ij}(\{\underline {p}\},\mu) 
{\boldsymbol {\mathcal S}}_{ij}(\vec{x}_\perp,\mu,\nu)\rangle\,.\nn
\end{align}
Here $d\Omega$ denotes the differential phase-space element of the final state with momenta $\{p_3, p_4, p_V\}$. The functions $B_{i/N_a}(z_a,\vec x_\perp,\mu,\nu)$ are the TMD beam functions, currently known to three loops~\cite{Luo:2019hmp,Luo:2019bmw,Luo:2019szz,Ebert:2020yqt,Luo:2020epw}, and encode collinear parton splittings in the incoming beams with transverse momentum $|\vec p_\perp|\sim |\vec x_\perp|^{-1}$ and momentum fraction $z_a$. The quantities $\boldsymbol{\mathcal H}_{ij}$ and $\boldsymbol{\mathcal S}_{ij}$ denote the hard Wilson coefficients and the soft function, respectively, both of which are matrices in color space. The soft function is currently available at one loop~\cite{Zhu:2012ts,Catani:2014qha,Shao:2025qgv}, and its average over the azimuthal angle of $\vec x_\perp$ is known numerically at two loops~\cite{Angeles-Martinez:2018mqh,Catani:2023tby,Devoto:2025eyc}; however, its full angular dependence is required, together with ongoing progress in the computation of three-loop soft anomalous dimensions~\cite{Liu:2022elt,Gardi:2025lws}, for evaluating Eq.~\eqref{eq:TMDfac} at next-to-next-to-next-to-leading logarithmic (N$^3$LL) accuracy, which is currently demanded by LHC measurements.
The scales $\mu$ and $\nu$ denote the \textit{renormalization} scales associated with ultraviolet and rapidity divergences~\cite{Collins:2003fm,Becher:2010tm,Becher:2011dz,Chiu:2011qc,Li:2016axz}, whose corresponding renormalization-group equations (RGEs) drive the resummation of the cross section.

A second important instance of factorization arises in the threshold regime of $Q\bar Q V$ production, defined by the limit $1 - M^2/\hat s \to 0$. In this region the cross section receives large logarithmic enhancements from soft-gluon emissions, and it satisfies the factorization formula~\cite{Ahrens:2010zv}
\begin{align}\label{eq:Thresfac}
&\frac{d\sigma}{d M^2 d\Omega} =\int \!\! dz_1 \int \!\!d z_2 
 \sum_{i,j} f_{i/N_1}(z_1,\mu) f_{j/N_2}(z_2,\mu) \\
&\times \int\!\! \frac{dx^0}{2\pi} e^{i \,{\sqrt {\hat s}}(1-\hat z)\, x^0} 
{\rm tr} \langle {\boldsymbol {\mathcal H}}_{ij}(\{\underline {p}\},\mu) 
{\boldsymbol {\mathcal S}}_{ij}(\{x^0, {\vec x}=0\},\mu)\rangle \,. \nn
\end{align}
Here, $f_{i/N}$ are the parton distribution functions (PDFs), $\hat z = M^2/\hat s$, and $\hat s = (z_1 P_1 + z_2 P_2)^2$. Unlike in the TMD factorization of Eq.~\eqref{eq:TMDfac}, the threshold factorization does not involve rapidity divergences, so only the renormalization scale $\mu$ appears. In this case the soft function is known at two loops for $Q\bar{Q}$ production~\cite{Czakon:2013hxa,Wang:2018vgu}, but not for $Q\bar{Q}V$.

Despite arising in seemingly distinct physical limits, the two factorization frameworks are closely connected: the respective soft functions correspond to the same underlying fully differential soft function ${\boldsymbol {\mathcal S}}(x)$, evaluated in different regions of coordinate space $x^\mu$. This connection provides a unified perspective on the resummation of logarithmic enhancements in both the small-$p_\perp$ and threshold regimes. ${\boldsymbol {\mathcal S}}(x)$ is defined as
\begin{align}\label{eq:sdef}
{\boldsymbol {\mathcal S}}(x)=
\frac{1}{d_R}\left \langle 0 \left | \overline{\bf T}\left[{\boldsymbol O}_s^\dagger (x)\right]
{\bf T} \left[{\boldsymbol O}_s(0) \right] \right | 0 \right \rangle \,,
\end{align}
where $d_R = N_c$ ($d_R = N_c^2 - 1$) for quarks (gluons) in the initial state. In the conventions of Ref.~\cite{Ahrens:2010zv}, the soft operator is
\begin{equation}
    \boldsymbol{O}_s(x) = Y_{n_1}(x)\,Y_{n_2}^\dagger(x)\,Y_{v_3}^\dagger(x)\,Y_{v_4}(x)\,,
\end{equation}
where $Y_{n_{1,2}}(x)$ and $Y_{v_{3,4}}(x)$ are soft Wilson lines along the directions of the massless partons $p_1, p_2$ and the heavy quarks $p_3, p_4$, respectively. The dependence of the soft functions in Eqs.~\eqref{eq:TMDfac} and \eqref{eq:Thresfac} on the incoming parton indices $i,j$ enters solely through the color representation of the corresponding Wilson lines, and it is omitted in the following to simplify the notation.

In the rapidity regularization scheme of Ref.~\cite{Li:2016ctv}, the TMD soft function in Eq.~\eqref{eq:TMDfac} corresponds to Eq.~\eqref{eq:sdef} evaluated at $x_\perp^\mu \to x^\mu = (i\tau, i\tau, \vec x_\perp)$, with $\tau = b_0/\nu$ and $b_0 = 2 e^{-\gamma_E}$, whereas the threshold soft function in Eq.~\eqref{eq:Thresfac} corresponds to Eq.~\eqref{eq:sdef} evaluated at $x^\mu = (x^0, \vec 0)$.

Beyond these examples, the fully differential soft function plays a central role in the computation of numerous collider observables (e.g.~\cite{Buonocore:2022mle,Abreu:2022sdc,Fu:2024fgj,vanBeekveld:2025zjh}) involving heavy-quark final states, highlighting its universality across a wide range of key collider physics applications.
In Ref.~\cite{Liu:2024hfa}, we presented the two-loop corrections to the fully-differential soft function for heavy-quark production at lepton colliders, while the two-loop massless counterpart was derived in Ref.~\cite{Li:2011zp}. All remaining contributions are computed in the present work, yielding novel analytic predictions for both the TMD and threshold soft functions in full kinematics.

\paragraph*{Soft function in momentum space.---}
We now turn to the calculation of the fully differential soft function through two loops for the process under consideration, retaining the full dependence on the generic kinematics of the scattering process. To streamline the computation, it is convenient to employ the following momentum-space representation of the soft function~\cite{Liu:2024hfa}
\begin{equation}\label{eq:sdef-mom}
{\boldsymbol {\mathcal S}}(x)=\int d\omega \,e^{i\omega t}
{\boldsymbol S}(\omega,\{\eta,\underline{n},\underline{v}\})\,,
\end{equation}
where
\begin{align}
{\boldsymbol S}(\omega,&\{\eta,\underline{n},\underline{v}\})=\\&\times\frac{1}{d_R}\left \langle 0 \left | \overline{\bf T}\left[{\boldsymbol O}_s^\dagger (0)\right]\delta(\omega-\eta\cdot{\hat p})
{\bf T} \left[{\boldsymbol O}_s(0) \right] \right | 0 \right \rangle\,.\nn
\end{align}
Here, $\hat p^\mu$ is an operator selecting the total momentum of the soft radiation in the final state, and $\eta^\mu$ is a dimensionless four-vector aligned with $x^\mu$, such that $x^\mu = t\,\eta^\mu$. The TMD and threshold soft functions of Eqs.~\eqref{eq:TMDfac} and \eqref{eq:Thresfac} are then obtained through the simple substitutions~\cite{Liu:2024hfa}
\begin{align}
\eta^\mu &= (\tau/|\vec x_\perp|, -i \cos\phi, -i \sin\phi, 0), \quad t = i |\vec x_\perp|, \label{eq:tmd-case} \\
\eta^\mu &= (1, 0, 0, 0), \quad t = x^0, \label{eq:threshold-case}
\end{align}
where $\phi$ is the azimuthal angle between $\vec p_\perp$ and $\vec x_\perp$.

\begin{figure*}
	\begin{center}
		\subfigure[\,massless-massless dipole]{\label{fig:rr12}\includegraphics[width=0.21\textwidth]{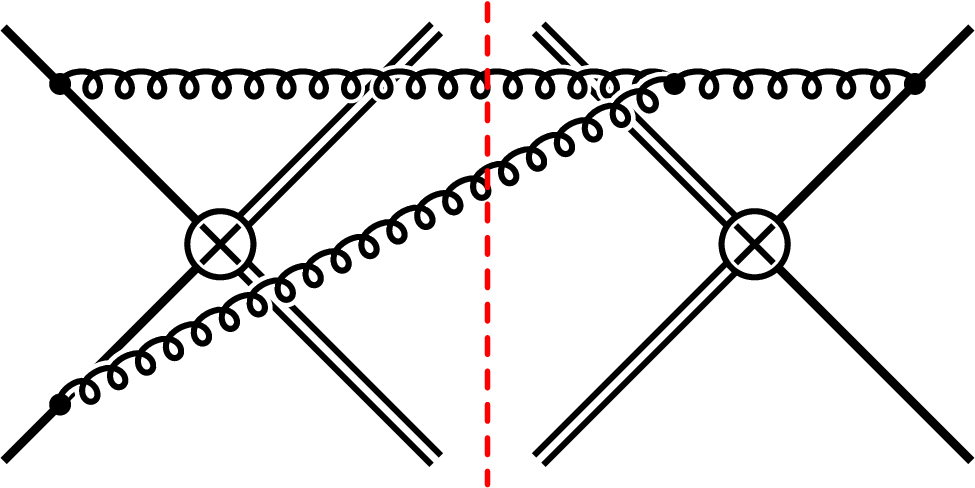}}\hspace{0.6cm}
		\subfigure[\,massless-massive dipole]{\label{fig:rr13}\includegraphics[width=0.21\textwidth]{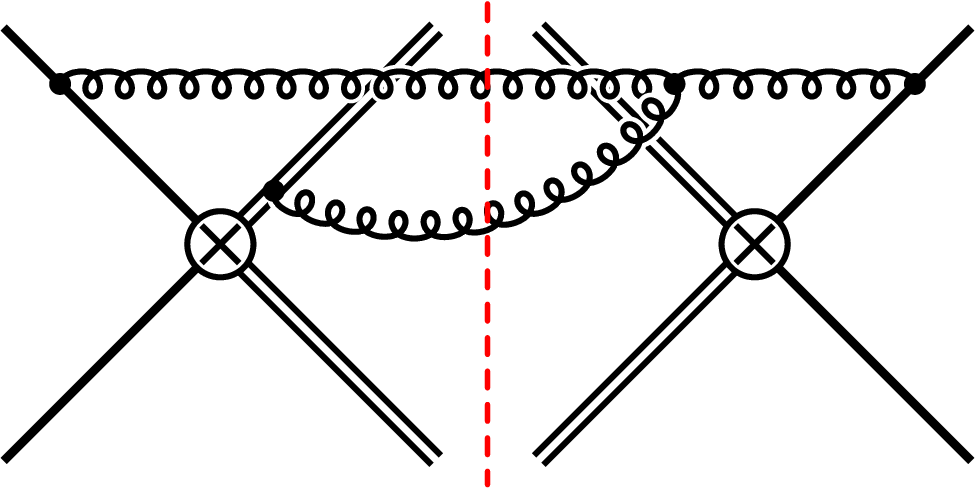}}\hspace{0.6cm}
		\subfigure[\,massive-massive dipole]{\label{fig:rr34}\includegraphics[width=0.21\textwidth]{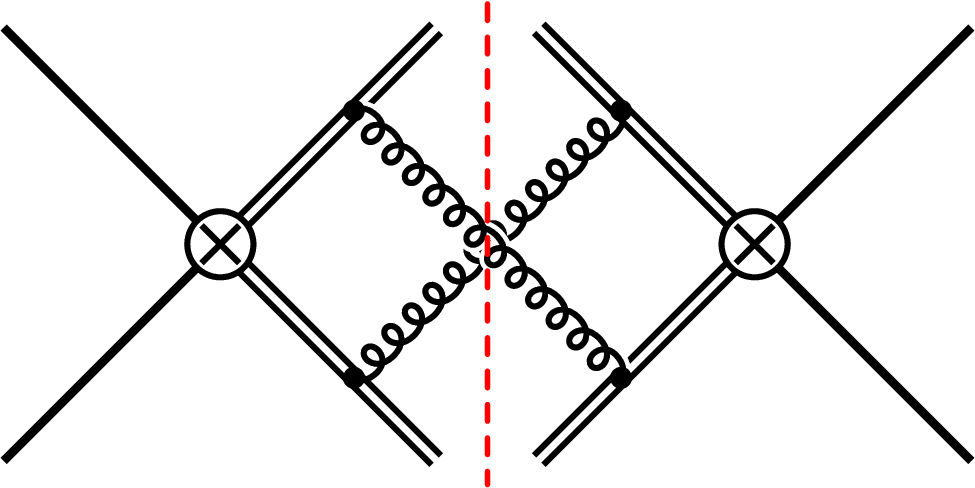}}\hspace{0.6cm}
        \subfigure[\, $(\bmT_1 \cdot \bmT_2)(\bmT_3 \cdot \bmT_4)$]{\label{fig:rr1234}\includegraphics[width=0.21\textwidth]{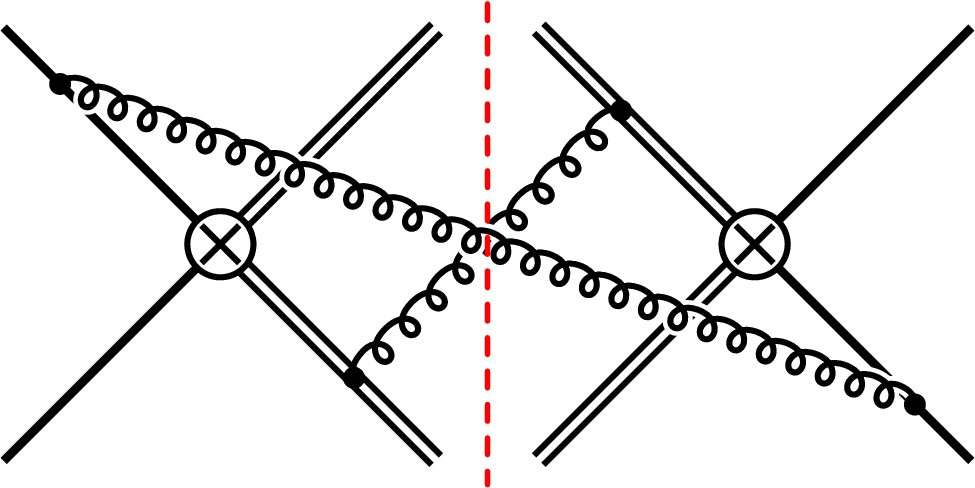}}
        \\
    	\subfigure[\, ${\boldsymbol \cT}_{123}$ tripole diagram]{\label{fig:vr123}\includegraphics[width=0.21\textwidth]{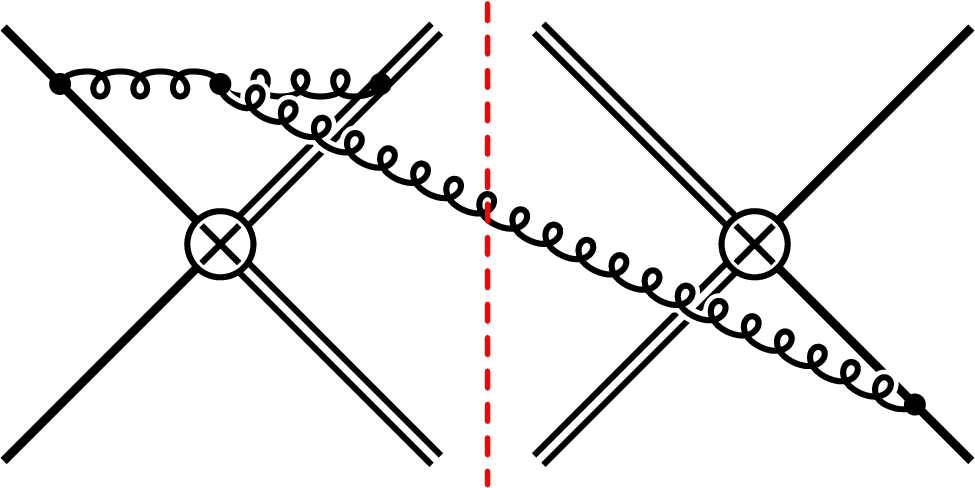}}\hspace{0.6cm}
		\subfigure[\, ${\boldsymbol \cT}_{134}$ tripole diagram]{\label{fig:vr134}\includegraphics[width=0.21\textwidth]{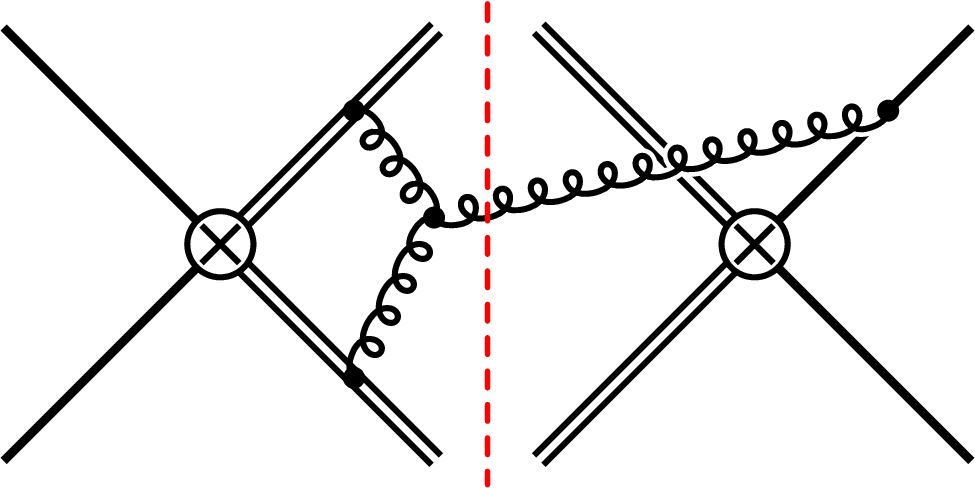}}\hspace{0.6cm}
		\subfigure[\, $\bmT^a_1 \bmT^b_4 \{\bmT^a_3,\bmT^b_3\}$]{\label{fig:rr1334}\includegraphics[width=0.21\textwidth]{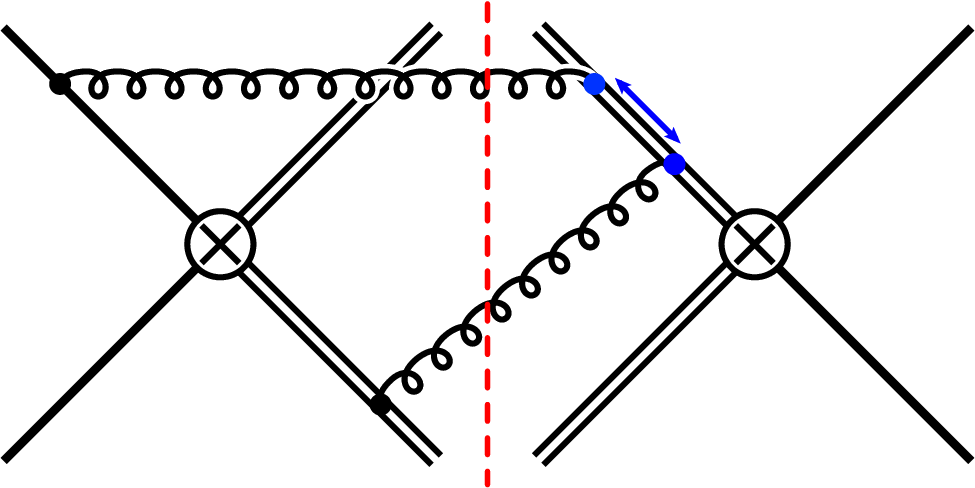}}\hspace{0.6cm}
        \subfigure[\, RR tripole cut]{\label{fig:rr134}\includegraphics[width=0.21\textwidth]{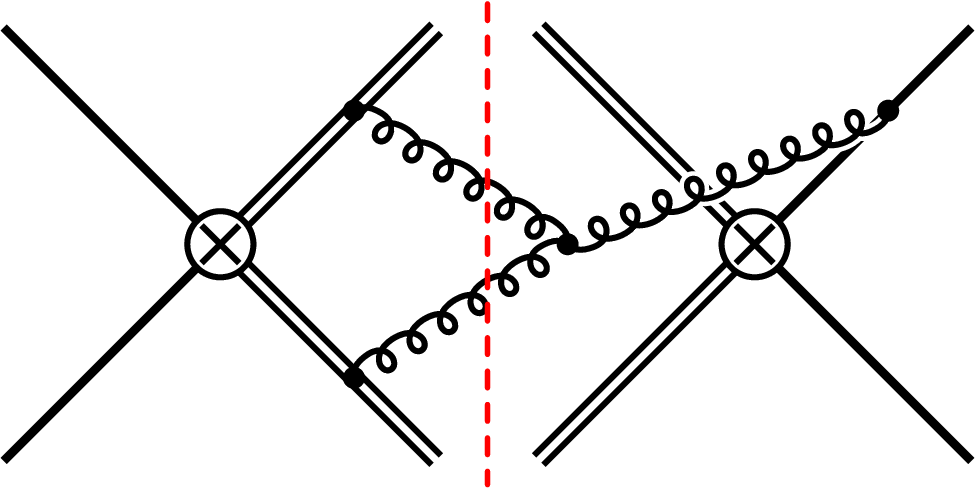}}
	\end{center}
	\caption{\label{fig:feyndia} Examples for two-loop Feynman diagrams that contribute to the soft function.}
\end{figure*}

Since $\boldsymbol S$ is a matrix in color space, we express it in terms of color generators $\bmT_i$~\cite{Catani:1996jh,Catani:1996vz}. The bare soft function in momentum space admits a perturbative expansion in the strong coupling $\alpha_s$
\begin{align}
{\boldsymbol S}&(\omega,\{\eta,\underline{n},\underline{v}\};\epsilon)=\delta(\omega)\, {\boldsymbol 1}
 \nn\\
 &+ \sum_{n=1}\frac{1}{\omega}\left(\frac{\eta^2\mu^2}{4\,\omega^2}\right)^{n\epsilon}
 \left(\frac{Z_\alpha\alpha_s}{4\pi}\right)^n
 {\boldsymbol S}^{(n)}\left(\chi,\rho,\underline{t},\underline{\kappa};\epsilon\right)\,.
 \label{eq:sexp}
\end{align}
The $n$-th order coefficient, $\boldsymbol S^{(n)}$, depends only on the cross ratios of the four vectors $n_{1,2}$ and $v_{3,4}$, a consequence of the rescaling invariance of the Wilson lines in $\boldsymbol O_s$. These cross ratios involve either the directions of the massive partons~\cite{Liu:2024hfa}
\begin{gather}\label{eq:xratios34}
\rho=\frac{v_3\cdot v_4}{\sqrt{v_3^2}\sqrt{v_4^2}}\,, \quad
\kappa_{I}=\frac{v_I\cdot \eta}{\sqrt{v_I^2}\sqrt{\eta^2}} \quad 
\mbox{with } \ I=3,4 \,,
\end{gather}
or those of both the massive and massless ones
\begin{gather}\label{eq:xratios1234}
\chi=\frac{2(\eta\cdot n_1)(\eta\cdot n_2)}{\eta^2 (n_1\cdot n_2)}\,,\quad
t_{jI}=\frac{\sqrt{v_I^2}(\eta\cdot n_j)}{\sqrt{\eta^2}(n_j\cdot v_I)}\,,
\end{gather}
with $j=1,2$ and $I=3,4$\,. 

At one loop, the soft function receives contributions only from a single real emission, which connects either two Wilson lines (a dipole correlator, $\bmT_i\cdot\bmT_j$) or a single massive Wilson line. At two loops, both virtual-real (VR) and double-real (RR) corrections contribute to $\boldsymbol S^{(2)}$. In addition to two-loop corrections to dipole correlators, diagrams in which soft radiation connects three different Wilson lines (tripole correlators, $\boldsymbol{\mathcal T}_{ijk} = i f^{abc} \bmT_i^a \bmT_j^b \bmT_k^c$) must be taken into account.

Representative Feynman diagrams with tripole correlators are shown in Figs.~\ref{fig:vr123} and~\ref{fig:vr134}, along with their color structures. Tripole contributions appear at two loops exclusively through VR corrections, since the corresponding RR contribution (e.g., through diagrams like that of Fig.~\ref{fig:rr134}) vanish due to charge conjugation, $(\boldsymbol{\mathcal T}_{ijk})^\dagger = -\boldsymbol{\mathcal T}_{ijk}$. Consequently, the sum of all RR diagrams in which soft gluons connect three different Wilson lines can be expressed as a sum of products of two one-loop integrals after partial fractioning, with the color structure $\bmT_i^a \bmT_j^b \{\bmT_k^a, \bmT_k^b\}$. Double-dipole diagrams, such as that shown in Fig.~\ref{fig:rr1234}, contribute with color structures $(\bmT_i^a \bmT_j^a)(\bmT_k^b \bmT_l^b)$ and can similarly be factorized into products of one-loop integrals.

In terms of analytic complexity, massless-massless dipoles depend only on the variable $\chi$, while massless-massive $n_j\!-\!v_I$ dipoles depend on two variables, $t_{jI}$ and $\kappa_I$. Tripole correlators $\boldsymbol{\mathcal T}_{12I}$ and $\boldsymbol{\mathcal T}_{j34}$ depend on $\{\chi, t_{1I}, t_{2I}, \kappa_I\}$ and $\{t_{j3}, t_{j4}, \rho, \kappa_3, \kappa_4\}$, respectively, rendering their calculation highly nontrivial.

\paragraph*{Computational method.---}
Although the soft function can in principle be obtained via phase-space integrals of the soft currents~\cite{Catani:1999ss,Catani:2000pi,Czakon:2011ve,Bierenbaum:2011gg,Czakon:2018iev,Angeles-Martinez:2018mqh}, performing analytical integrations for massive dipole and tripole correlations is extremely challenging. We therefore employ modern multi-loop techniques to organize the calculation.

Amplitudes are generated with \texttt{qgraf} and, after Dirac and Lorentz algebra, are expressed in terms of scalar Feynman integrals using reverse unitarity~\cite{Anastasiou:2002yz,Anastasiou:2003yy,Anastasiou:2003ds}, which replaces the $\delta(\omega-\eta\cdot \hat p)$ function in Eq.~\eqref{eq:sdef} by a Cutkosky cut. Integration-by-parts (IBP) reduction is then performed with \texttt{FIRE6}~\cite{Smirnov:2019qkx} and \texttt{Kira}~\cite{Klappert:2020nbg} to express the soft function as a linear combination of master integrals (MIs). These MIs are evaluated via differential equations (DEs)~\cite{Kotikov:1990kg,Kotikov:1991hm,Kotikov:1991pm,Bern:1992em,Bern:1993kr,Remiddi:1997ny,Gehrmann:1999as}, with the DEs derived using \texttt{LiteRed}~\cite{Lee:2013mka}. The complete basis consists of 9, 30, and 51 MIs for the VR corrections to $n_i\!-\!v_I$, $n_1\!-\!n_2\!-\!v_I$, and $n_j\!-\!v_3\!-\!v_4$ correlations, respectively, and 20 MIs for the RR corrections to $n_i\!-\!v_I$ correlations.

The MIs are evaluated analytically by solving the DEs in its canonical form~\cite{Henn:2013pwa}, obtained by rotating the MI basis into a set of uniform transcendental (UT) weight integrals and applying Magnus and Dyson series expansions~\cite{Argeri:2014qva}. Candidate UT integrals are identified via their leading singularities~\cite{Cachazo:2008vp,Arkani-Hamed:2010pyv}.

The DEs corresponding to the tripole correlators $\boldsymbol{\mathcal T}_{12I}$ and $\boldsymbol{\mathcal T}_{j34}$ contain two ($R^A_{1,2}$) and four ($R^B_{1,\dots,4}$) square roots, respectively, whose explicit expressions are given in Eq.~\eqref{eq:srootsA},~\eqref{eq:srootsB} of the supplemental material to this Letter. The boundary conditions are evaluated at the phase-space point $\eta^\mu = v_3^\mu = v_4^\mu$ with the two light-like vectors $n_1$ and $n_2$ being back-to-back, i.e., $\chi = \rho = t_{jI} = \kappa_I = 1$. For the tripole correlator $\boldsymbol{\mathcal T}_{j34}$, these contain the Coulomb singularities characteristic of heavy-quark production. All MIs are then checked numerically with \texttt{AMFlow}~\cite{Liu:2017jxz,Liu:2022chg} at random benchmark points.
%configurations that  singular terms $(\rho^2-1)^\epsilon$ arising from Coulomb singularities are also included in the boundary conditions.

With the canonical DEs and these boundary conditions, the MIs can be solved iteratively order by order in $\epsilon$. In practice, the solution is subtle due to the continuation to different regions of the coordinate $x^\mu$. For timelike $x^\mu$ (relevant for threshold kinematics), the MI evolution from the boundary is path independent, as a consequence of the fact that the boundary conditions are set at a time-like point. However, for spacelike $x^\mu$ (relevant for the TMD soft function), the evolution of the tripole contribution crosses branch cuts originating from $n_i\cdot \eta = 0$, requiring a careful analytical continuation. To resolve it, we perform the variable changes $\{t_{1I}, t_{2I}\} = \{\sqrt{\chi}\, t_{12I} w_I, \sqrt{\chi}\, t_{12I}/w_I\}$ and $\{t_{j3}, t_{j4}\} = \{\xi z_j, \xi / z_j\}$ for $\boldsymbol{\mathcal T}_{12I}$ and $\boldsymbol{\mathcal T}_{j34}$, respectively, where we express $\chi$ and $\xi$ as
\begin{gather}
\chi = \frac{\tau^2}{|\vec{x}_\perp|^2}, \qquad
\xi = \frac{\tau^2}{|\vec{x}_\perp|^2} \frac{\sqrt{v_3^2\, v_4^2}}{(n_j\cdot v_3)(n_j\cdot v_4)}\,.\label{eq:TMDchixi}
\end{gather}
In this way, the dependence on $n_i\cdot \eta$ is fully absorbed into $\chi$ and $\xi$, which relate to the rapidity regulator $\tau$ in the TMD soft function. We first evolve the MIs in $\chi$ ($\xi$) from the boundary to the values in Eq.~(\ref{eq:TMDchixi}). We then expand around $\tau\to 0$, resolving the logarithmic dependence on $\tau$ via asymptotic expansion. Finally, the remaining variables are then evolved by setting $\chi = 0$ ($\xi = 0$), hence taking care of the correct continuation and yielding a path-independent evolution.

To express the integrals in terms of GPLs, the square roots in the DEs must be rationalized. Using variable changes and \texttt{RationalizeRoots}~\cite{Besier:2019kco}, we rationalize the DEs for timelike and spacelike measurements individually. For tripole contributions, the rationalization is performed after the evolution and asymptotic expansion in $\chi$ ($\xi$). The resulting DEs for the MIs vector $\vec{I}(\{\underline s\},\epsilon)$ can be then written in $d\log$ form
\begin{align}
d \vec{I} (\{\underline s\},\epsilon) = \epsilon \left( \sum_{k=1}^{N_l} c_k \, d \log W_k(\{\underline s\}) \right) \vec{I}(\{\underline s\},\epsilon),
\end{align}
where $c_k$ are rational-number matrices, $W_k$ are symbol letters expressed as polynomials of the kinematic variables $\{\underline s\}$, and $N_l$ denotes the number of letters. For the two-loop corrections to the $n_j\!-\!v_I$, $n_1\!-\!n_2\!-\!v_I$, and $n_j\!-\!v_3\!-\!v_4$ correlations, the DE systems involve 7, 22, and 39 symbol letters, respectively.

\paragraph*{Result and consistency checks.---}
With the analytical results for the MIs, we obtain the bare soft function through two loops in momentum space. The corresponding position-space soft function is obtained by inverting the one-dimensional Fourier transform in Eq.~\eqref{eq:sdef-mom}, which affects only the $\omega$-dependent pre-factor in Eq.~\eqref{eq:sexp}:
\begin{align}
\int d\omega\, e^{i \omega t}\frac{1}{\omega}\left(\frac{\eta^2\mu^2}{4\,\omega^2}\right)^{n\epsilon} =
e^{2n\epsilon L} e^{-2n\epsilon \gamma_E}\Gamma(-2n\epsilon),
\end{align}
where $L = \ln\left(-i x_0 \mu/b_0\right)$ for the threshold soft function and $L = \ln\left(x_\perp \mu/b_0\right)$ for the TMD soft function.\footnote{We note that, the threshold soft function is often presented in Laplace space instead of Fourier space. The relation between the two spaces simply amounts to a redefinition of the logarithm $L$.}

The UV divergences of the bare soft function are removed via a local multiplicative renormalization,
$\boldsymbol{\mathcal S}(x,\mu) = {\boldsymbol Z}_S^\dagger(\mu)\, \boldsymbol{\mathcal S}(x;\epsilon)\, {\boldsymbol Z}_S(\mu)$,
where ${\boldsymbol Z}_S$ is a matrix in color space. The renormalization constant is determined by the soft anomalous dimension ${\bm \Gamma}_S$, which governs the infrared divergences of scattering amplitudes with massive partons~\cite{Becher:2009cu,Becher:2009qa}:
\begin{align}
{\boldsymbol Z}_S (\mu) = {\cal P} \exp \int_\mu^\infty \frac{d\bar \mu}{\bar \mu} {\boldsymbol \Gamma}_S(\alpha_s(\bar \mu), \bar \mu)\,.
\end{align}
The soft anomalous dimension takes the form~\cite{Becher:2009kw,Ferroglia:2009ep,Kidonakis:2009ev,Grozin:2014hna}
\begin{align}
{\boldsymbol \Gamma}_S & = \bmT_1\cdot\bmT_2\, \gcusp(\alpha_s)\left(2L + \ln \chi + i\pi\right) 
\nn\\
&+ \sum_{I,j} \bmT_I\cdot\bmT_j \,\gcusp(\alpha_s)\left(L + \ln t_{jI}\right)
\nn\\
&- \bmT_3 \cdot \bmT_4\, \gcusp(\rho,\alpha_s)
+ \gamma^S(\alpha_s) {\boldsymbol 1} \nn\\
&+ \sum_{j=1,2} {\boldsymbol \cT}_{j34} \frac{\alpha_s}{2\pi} g(\rho)\, \gcusp(\alpha_s) \ln \frac{t_{j3}}{t_{j4}}
+ \mathcal O(\alpha_s^3)\,.
\end{align}
Here, $\gcusp(\alpha_s)$ and $\gcusp(\rho,\alpha_s)$ denote the light-like and angle-dependent cusp anomalous dimensions, respectively, and $\gamma^S = \gamma^{\phi_a} + \gamma^{\phi_b} + \gamma^a + \gamma^b + 2 \gamma^Q$, where $\gamma^{\phi_{a,b}}$ are the PDF anomalous dimensions, and $\gamma^{a,b,Q}$ those of the initial-state partons and heavy quarks. Finally, the function $g(\rho)$ is given by~\cite{Ferroglia:2009ii}
\begin{align}
\frac{g(\rho)}{i\pi} =  \frac{-\rho}{\sqrt{\rho^2-1}} \ln\bigl(4(\rho^2-1)\bigr) + 2 \ln\Bigl(\rho + \sqrt{\rho^2-1}\Bigr).\nn
\end{align}

%---
The cancellation of UV poles via this renormalization procedure provides a stringent consistency check of our calculation. We further verify that the TMD soft function satisfies the rapidity RGE~\cite{Li:2016axz}:
%\begin{align}
%\frac{d\,\ln \boldsymbol{\mathcal S}(x_\perp,\mu,\nu)}{d \ln \nu} = 
%4 \int_\mu^{b_0/|\vec{x}_\perp|} &\frac{d\bar \mu}{\bar \mu} \gcusp(\alpha_s(\bar \mu)) 
%\\&+ 2 \, \gamma^R\Big(\alpha_s(b_0/|\vec{x}_\perp|)\Big), \nn\label{eq:sRRGE}
%\end{align}
\begin{align}
\frac{d\,\boldsymbol{\mathcal S}(\vec{x}_\perp,\mu,\nu)}{d \ln \nu^2} &= 
\bigg[ \int_{\mu^2}^{b^2_0/|\vec{x}_\perp|^2} \frac{d\bar \mu^2}{\bar \mu^2} \gcusp(\alpha_s(\bar \mu)) 
\\&+\gamma^R\left(\alpha_s(b_0/|\vec{x}_\perp|)\right)\bigg]\boldsymbol{\mathcal S}(\vec{x}_\perp,\mu,\nu), \nn\label{eq:sRRGE}
\end{align}
where $\gamma^R$ is the rapidity anomalous dimension. This check is highly nontrivial: although the tripole correlators $\boldsymbol{\mathcal T}_{12I}$ and $\boldsymbol{\mathcal T}_{j34}$ individually depend on $\nu$, their combination becomes $\nu$-independent once color conservation, $\sum_{i=1}^4 \bmT_i = 0$, is employed.

The final two-loop renormalized soft functions for the TMD and threshold factorization theorems are provided in Ref.~\cite{zenodo} as a function of over 8000 GPLs, encoding the full color and kinematic dependence of the process.

\paragraph*{Outlook.---}
In this Letter, we have presented the first complete analytic two-loop calculation of the fully-differential soft function relevant for heavy-quark pair production in association with a color-singlet system at hadron colliders. Derived within the framework of soft-collinear effective theory, our result represents a crucial building block for precision predictions of scattering observables that probe the properties of the heavy quarks at the LHC. The decomposition of the soft function into dipole and tripole color correlators, together with the full dependence on the directions of all the Wilson lines, further provides essential ingredients for processes involving additional QCD jets, both at lepton and hadron colliders, and offers deep insight into the universal infrared structure of QCD amplitudes with massive quarks.
Our analytic results are made publicly available at~\cite{zenodo}. Looking ahead, a natural next step is to exploit these results for high-precision collider phenomenology, enabling both fixed-order and resummed predictions for heavy-quark processes, as well as NNLO-accurate Monte Carlo simulations for the LHC. Furthermore, the methods developed here open the door to analogous calculations for processes with heavy quarks and additional jets, promising to deepen our understanding of QCD dynamics in the multi-scale regime and to enhance the theoretical toolkit for precision collider physics.

\vspace{1mm}
\paragraph*{Acknowledgments.---}
Z.L.L. is grateful to Ding Yu Shao, Hua-Sheng Shao, Guo-Xing Wang, Xiao-Feng Xu, Li Lin Yang and Yang Zhang for stimulating discussions. We thank the Mainz Institute of Theoretical Physics (MITP) for hospitality during the final stages of this work. 
We are also grateful to Luca Buonocore for discussions and ongoing collaboration in related projects.
Z.L.L. is supported by Institute of High Energy Physics (IHEP, CAS) under Grant No. E35159U1. The work of P.M. and Z.L.L. was funded by the European Union (ERC, grant agreement No. 101044599, JANUS). Views and opinions
expressed are however those of the authors only and do not necessarily reflect those of the European Union or the European Research Council Executive Agency. Neither the European Union nor the granting authority can be held responsible for them.
%Z.L.L thanks Robin Br\"user for providing the codes to perform the topology mapping and partial fractioning. 

\bibliographystyle{apsrev4-1}
\bibliography{letter}

%merlin.mbs apsrev4-1.bst 2010-07-25 4.21a (PWD, AO, DPC) hacked
%Control: key (0)
%Control: author (72) initials jnrlst
%Control: editor formatted (1) identically to author
%Control: production of article title (-1) disabled
%Control: page (0) single
%Control: year (1) truncated
%Control: production of eprint (0) enabled
\begin{thebibliography}{94}%
\makeatletter
\providecommand \@ifxundefined [1]{%
 \@ifx{#1\undefined}
}%
\providecommand \@ifnum [1]{%
 \ifnum #1\expandafter \@firstoftwo
 \else \expandafter \@secondoftwo
 \fi
}%
\providecommand \@ifx [1]{%
 \ifx #1\expandafter \@firstoftwo
 \else \expandafter \@secondoftwo
 \fi
}%
\providecommand \natexlab [1]{#1}%
\providecommand \enquote  [1]{``#1''}%
\providecommand \bibnamefont  [1]{#1}%
\providecommand \bibfnamefont [1]{#1}%
\providecommand \citenamefont [1]{#1}%
\providecommand \href@noop [0]{\@secondoftwo}%
\providecommand \href [0]{\begingroup \@sanitize@url \@href}%
\providecommand \@href[1]{\@@startlink{#1}\@@href}%
\providecommand \@@href[1]{\endgroup#1\@@endlink}%
\providecommand \@sanitize@url [0]{\catcode `\\12\catcode `\$12\catcode
  `\&12\catcode `\#12\catcode `\^12\catcode `\_12\catcode `\%12\relax}%
\providecommand \@@startlink[1]{}%
\providecommand \@@endlink[0]{}%
\providecommand \url  [0]{\begingroup\@sanitize@url \@url }%
\providecommand \@url [1]{\endgroup\@href {#1}{\urlprefix }}%
\providecommand \urlprefix  [0]{URL }%
\providecommand \Eprint [0]{\href }%
\providecommand \doibase [0]{http://dx.doi.org/}%
\providecommand \selectlanguage [0]{\@gobble}%
\providecommand \bibinfo  [0]{\@secondoftwo}%
\providecommand \bibfield  [0]{\@secondoftwo}%
\providecommand \translation [1]{[#1]}%
\providecommand \BibitemOpen [0]{}%
\providecommand \bibitemStop [0]{}%
\providecommand \bibitemNoStop [0]{.\EOS\space}%
\providecommand \EOS [0]{\spacefactor3000\relax}%
\providecommand \BibitemShut  [1]{\csname bibitem#1\endcsname}%
\let\auto@bib@innerbib\@empty
%</preamble>
\bibitem [{\citenamefont {Janot}(2015)}]{Janot:2015yza}%
  \BibitemOpen
  \bibfield  {author} {\bibinfo {author} {\bibfnamefont {P.}~\bibnamefont
  {Janot}},\ }\href {\doibase 10.1007/JHEP04(2015)182} {\bibfield  {journal}
  {\bibinfo  {journal} {JHEP}\ }\textbf {\bibinfo {volume} {04}},\ \bibinfo
  {pages} {182} (\bibinfo {year} {2015})},\ \Eprint
  {http://arxiv.org/abs/1503.01325} {arXiv:1503.01325 [hep-ph]} \BibitemShut
  {NoStop}%
\bibitem [{\citenamefont {Vos}\ \emph {et~al.}(2016)\citenamefont {Vos} \emph
  {et~al.}}]{Vos:2016til}%
  \BibitemOpen
  \bibfield  {author} {\bibinfo {author} {\bibfnamefont {M.}~\bibnamefont
  {Vos}} \emph {et~al.},\ }\href@noop {} {\  (\bibinfo {year} {2016})},\
  \Eprint {http://arxiv.org/abs/1604.08122} {arXiv:1604.08122 [hep-ex]}
  \BibitemShut {NoStop}%
\bibitem [{\citenamefont {Tumasyan}\ \emph {et~al.}(2021)\citenamefont
  {Tumasyan} \emph {et~al.}}]{CMS:2021klw}%
  \BibitemOpen
  \bibfield  {author} {\bibinfo {author} {\bibfnamefont {A.}~\bibnamefont
  {Tumasyan}} \emph {et~al.} (\bibinfo {collaboration} {CMS}),\ }\href
  {\doibase 10.1007/JHEP12(2021)180} {\bibfield  {journal} {\bibinfo  {journal}
  {JHEP}\ }\textbf {\bibinfo {volume} {12}},\ \bibinfo {pages} {180} (\bibinfo
  {year} {2021})},\ \Eprint {http://arxiv.org/abs/2107.01508} {arXiv:2107.01508
  [hep-ex]} \BibitemShut {NoStop}%
\bibitem [{\citenamefont {Tumasyan}\ \emph {et~al.}(2023)\citenamefont
  {Tumasyan} \emph {et~al.}}]{CMS:2022tkv}%
  \BibitemOpen
  \bibfield  {author} {\bibinfo {author} {\bibfnamefont {A.}~\bibnamefont
  {Tumasyan}} \emph {et~al.} (\bibinfo {collaboration} {CMS}),\ }\href
  {\doibase 10.1007/JHEP07(2023)219} {\bibfield  {journal} {\bibinfo  {journal}
  {JHEP}\ }\textbf {\bibinfo {volume} {07}},\ \bibinfo {pages} {219} (\bibinfo
  {year} {2023})},\ \Eprint {http://arxiv.org/abs/2208.06485} {arXiv:2208.06485
  [hep-ex]} \BibitemShut {NoStop}%
\bibitem [{\citenamefont {Aad}\ \emph {et~al.}(2024{\natexlab{a}})\citenamefont
  {Aad} \emph {et~al.}}]{ATLAS:2023eld}%
  \BibitemOpen
  \bibfield  {author} {\bibinfo {author} {\bibfnamefont {G.}~\bibnamefont
  {Aad}} \emph {et~al.} (\bibinfo {collaboration} {ATLAS}),\ }\href {\doibase
  10.1007/JHEP07(2024)163} {\bibfield  {journal} {\bibinfo  {journal} {JHEP}\
  }\textbf {\bibinfo {volume} {07}},\ \bibinfo {pages} {163} (\bibinfo {year}
  {2024}{\natexlab{a}})},\ \Eprint {http://arxiv.org/abs/2312.04450}
  {arXiv:2312.04450 [hep-ex]} \BibitemShut {NoStop}%
\bibitem [{\citenamefont {Hayrapetyan}\ \emph
  {et~al.}(2025{\natexlab{a}})\citenamefont {Hayrapetyan} \emph
  {et~al.}}]{CMS:2024fdo}%
  \BibitemOpen
  \bibfield  {author} {\bibinfo {author} {\bibfnamefont {A.}~\bibnamefont
  {Hayrapetyan}} \emph {et~al.} (\bibinfo {collaboration} {CMS}),\ }\href
  {\doibase 10.1007/JHEP02(2025)097} {\bibfield  {journal} {\bibinfo  {journal}
  {JHEP}\ }\textbf {\bibinfo {volume} {02}},\ \bibinfo {pages} {097} (\bibinfo
  {year} {2025}{\natexlab{a}})},\ \Eprint {http://arxiv.org/abs/2407.10896}
  {arXiv:2407.10896 [hep-ex]} \BibitemShut {NoStop}%
\bibitem [{\citenamefont {Aad}\ \emph {et~al.}(2025)\citenamefont {Aad} \emph
  {et~al.}}]{ATLAS:2024gth}%
  \BibitemOpen
  \bibfield  {author} {\bibinfo {author} {\bibfnamefont {G.}~\bibnamefont
  {Aad}} \emph {et~al.} (\bibinfo {collaboration} {ATLAS}),\ }\href {\doibase
  10.1140/epjc/s10052-025-13740-x} {\bibfield  {journal} {\bibinfo  {journal}
  {Eur. Phys. J. C}\ }\textbf {\bibinfo {volume} {85}},\ \bibinfo {pages} {210}
  (\bibinfo {year} {2025})},\ \Eprint {http://arxiv.org/abs/2407.10904}
  {arXiv:2407.10904 [hep-ex]} \BibitemShut {NoStop}%
\bibitem [{\citenamefont {Hayrapetyan}\ \emph
  {et~al.}(2025{\natexlab{b}})\citenamefont {Hayrapetyan} \emph
  {et~al.}}]{CMS:2024mke}%
  \BibitemOpen
  \bibfield  {author} {\bibinfo {author} {\bibfnamefont {A.}~\bibnamefont
  {Hayrapetyan}} \emph {et~al.} (\bibinfo {collaboration} {CMS}),\ }\href
  {\doibase 10.1007/JHEP02(2025)177} {\bibfield  {journal} {\bibinfo  {journal}
  {JHEP}\ }\textbf {\bibinfo {volume} {02}},\ \bibinfo {pages} {177} (\bibinfo
  {year} {2025}{\natexlab{b}})},\ \Eprint {http://arxiv.org/abs/2410.23475}
  {arXiv:2410.23475 [hep-ex]} \BibitemShut {NoStop}%
\bibitem [{\citenamefont {Aad}\ \emph {et~al.}(2024{\natexlab{b}})\citenamefont
  {Aad} \emph {et~al.}}]{ATLAS:2024moy}%
  \BibitemOpen
  \bibfield  {author} {\bibinfo {author} {\bibfnamefont {G.}~\bibnamefont
  {Aad}} \emph {et~al.} (\bibinfo {collaboration} {ATLAS}),\ }\href {\doibase
  10.1007/JHEP05(2024)131} {\bibfield  {journal} {\bibinfo  {journal} {JHEP}\
  }\textbf {\bibinfo {volume} {05}},\ \bibinfo {pages} {131} (\bibinfo {year}
  {2024}{\natexlab{b}})},\ \Eprint {http://arxiv.org/abs/2401.05299}
  {arXiv:2401.05299 [hep-ex]} \BibitemShut {NoStop}%
\bibitem [{\citenamefont {Benedikt}\ \emph {et~al.}(2025)\citenamefont
  {Benedikt} \emph {et~al.}}]{FCC:2025lpp}%
  \BibitemOpen
  \bibfield  {author} {\bibinfo {author} {\bibfnamefont {M.}~\bibnamefont
  {Benedikt}} \emph {et~al.} (\bibinfo {collaboration} {FCC}),\ }\href@noop {}
  {\  (\bibinfo {year} {2025})},\ \Eprint {http://arxiv.org/abs/2505.00272}
  {arXiv:2505.00272 [hep-ex]} \BibitemShut {NoStop}%
\bibitem [{\citenamefont {von Manteuffel}\ \emph {et~al.}(2015)\citenamefont
  {von Manteuffel}, \citenamefont {Schabinger},\ and\ \citenamefont
  {Zhu}}]{vonManteuffel:2014mva}%
  \BibitemOpen
  \bibfield  {author} {\bibinfo {author} {\bibfnamefont {A.}~\bibnamefont {von
  Manteuffel}}, \bibinfo {author} {\bibfnamefont {R.~M.}\ \bibnamefont
  {Schabinger}}, \ and\ \bibinfo {author} {\bibfnamefont {H.~X.}\ \bibnamefont
  {Zhu}},\ }\href {\doibase 10.1103/PhysRevD.92.045034} {\bibfield  {journal}
  {\bibinfo  {journal} {Phys. Rev. D}\ }\textbf {\bibinfo {volume} {92}},\
  \bibinfo {pages} {045034} (\bibinfo {year} {2015})},\ \Eprint
  {http://arxiv.org/abs/1408.5134} {arXiv:1408.5134 [hep-ph]} \BibitemShut
  {NoStop}%
\bibitem [{\citenamefont {Gao}\ and\ \citenamefont
  {Zhu}(2014{\natexlab{a}})}]{Gao:2014nva}%
  \BibitemOpen
  \bibfield  {author} {\bibinfo {author} {\bibfnamefont {J.}~\bibnamefont
  {Gao}}\ and\ \bibinfo {author} {\bibfnamefont {H.~X.}\ \bibnamefont {Zhu}},\
  }\href {\doibase 10.1103/PhysRevD.90.114022} {\bibfield  {journal} {\bibinfo
  {journal} {Phys. Rev. D}\ }\textbf {\bibinfo {volume} {90}},\ \bibinfo
  {pages} {114022} (\bibinfo {year} {2014}{\natexlab{a}})},\ \Eprint
  {http://arxiv.org/abs/1408.5150} {arXiv:1408.5150 [hep-ph]} \BibitemShut
  {NoStop}%
\bibitem [{\citenamefont {Gao}\ and\ \citenamefont
  {Zhu}(2014{\natexlab{b}})}]{Gao:2014eea}%
  \BibitemOpen
  \bibfield  {author} {\bibinfo {author} {\bibfnamefont {J.}~\bibnamefont
  {Gao}}\ and\ \bibinfo {author} {\bibfnamefont {H.~X.}\ \bibnamefont {Zhu}},\
  }\href {\doibase 10.1103/PhysRevLett.113.262001} {\bibfield  {journal}
  {\bibinfo  {journal} {Phys. Rev. Lett.}\ }\textbf {\bibinfo {volume} {113}},\
  \bibinfo {pages} {262001} (\bibinfo {year} {2014}{\natexlab{b}})},\ \Eprint
  {http://arxiv.org/abs/1410.3165} {arXiv:1410.3165 [hep-ph]} \BibitemShut
  {NoStop}%
\bibitem [{\citenamefont {Zhu}\ \emph {et~al.}(2013)\citenamefont {Zhu},
  \citenamefont {Li}, \citenamefont {Li}, \citenamefont {Shao},\ and\
  \citenamefont {Yang}}]{Zhu:2012ts}%
  \BibitemOpen
  \bibfield  {author} {\bibinfo {author} {\bibfnamefont {H.~X.}\ \bibnamefont
  {Zhu}}, \bibinfo {author} {\bibfnamefont {C.~S.}\ \bibnamefont {Li}},
  \bibinfo {author} {\bibfnamefont {H.~T.}\ \bibnamefont {Li}}, \bibinfo
  {author} {\bibfnamefont {D.~Y.}\ \bibnamefont {Shao}}, \ and\ \bibinfo
  {author} {\bibfnamefont {L.~L.}\ \bibnamefont {Yang}},\ }\href {\doibase
  10.1103/PhysRevLett.110.082001} {\bibfield  {journal} {\bibinfo  {journal}
  {Phys. Rev. Lett.}\ }\textbf {\bibinfo {volume} {110}},\ \bibinfo {pages}
  {082001} (\bibinfo {year} {2013})},\ \Eprint {http://arxiv.org/abs/1208.5774}
  {arXiv:1208.5774 [hep-ph]} \BibitemShut {NoStop}%
\bibitem [{\citenamefont {Li}\ \emph {et~al.}(2013)\citenamefont {Li},
  \citenamefont {Li}, \citenamefont {Shao}, \citenamefont {Yang},\ and\
  \citenamefont {Zhu}}]{Li:2013mia}%
  \BibitemOpen
  \bibfield  {author} {\bibinfo {author} {\bibfnamefont {H.~T.}\ \bibnamefont
  {Li}}, \bibinfo {author} {\bibfnamefont {C.~S.}\ \bibnamefont {Li}}, \bibinfo
  {author} {\bibfnamefont {D.~Y.}\ \bibnamefont {Shao}}, \bibinfo {author}
  {\bibfnamefont {L.~L.}\ \bibnamefont {Yang}}, \ and\ \bibinfo {author}
  {\bibfnamefont {H.~X.}\ \bibnamefont {Zhu}},\ }\href {\doibase
  10.1103/PhysRevD.88.074004} {\bibfield  {journal} {\bibinfo  {journal} {Phys.
  Rev. D}\ }\textbf {\bibinfo {volume} {88}},\ \bibinfo {pages} {074004}
  (\bibinfo {year} {2013})},\ \Eprint {http://arxiv.org/abs/1307.2464}
  {arXiv:1307.2464 [hep-ph]} \BibitemShut {NoStop}%
\bibitem [{\citenamefont {Catani}\ \emph {et~al.}(2014)\citenamefont {Catani},
  \citenamefont {Grazzini},\ and\ \citenamefont {Torre}}]{Catani:2014qha}%
  \BibitemOpen
  \bibfield  {author} {\bibinfo {author} {\bibfnamefont {S.}~\bibnamefont
  {Catani}}, \bibinfo {author} {\bibfnamefont {M.}~\bibnamefont {Grazzini}}, \
  and\ \bibinfo {author} {\bibfnamefont {A.}~\bibnamefont {Torre}},\ }\href
  {\doibase 10.1016/j.nuclphysb.2014.11.019} {\bibfield  {journal} {\bibinfo
  {journal} {Nucl. Phys. B}\ }\textbf {\bibinfo {volume} {890}},\ \bibinfo
  {pages} {518} (\bibinfo {year} {2014})},\ \Eprint
  {http://arxiv.org/abs/1408.4564} {arXiv:1408.4564 [hep-ph]} \BibitemShut
  {NoStop}%
\bibitem [{\citenamefont {Catani}\ \emph {et~al.}(2018)\citenamefont {Catani},
  \citenamefont {Grazzini},\ and\ \citenamefont {Sargsyan}}]{Catani:2018mei}%
  \BibitemOpen
  \bibfield  {author} {\bibinfo {author} {\bibfnamefont {S.}~\bibnamefont
  {Catani}}, \bibinfo {author} {\bibfnamefont {M.}~\bibnamefont {Grazzini}}, \
  and\ \bibinfo {author} {\bibfnamefont {H.}~\bibnamefont {Sargsyan}},\ }\href
  {\doibase 10.1007/JHEP11(2018)061} {\bibfield  {journal} {\bibinfo  {journal}
  {JHEP}\ }\textbf {\bibinfo {volume} {11}},\ \bibinfo {pages} {061} (\bibinfo
  {year} {2018})},\ \Eprint {http://arxiv.org/abs/1806.01601} {arXiv:1806.01601
  [hep-ph]} \BibitemShut {NoStop}%
\bibitem [{\citenamefont {Ju}\ and\ \citenamefont
  {Sch\"onherr}(2024)}]{Ju:2024xhd}%
  \BibitemOpen
  \bibfield  {author} {\bibinfo {author} {\bibfnamefont {W.-L.}\ \bibnamefont
  {Ju}}\ and\ \bibinfo {author} {\bibfnamefont {M.}~\bibnamefont
  {Sch\"onherr}},\ }\href@noop {} {\  (\bibinfo {year} {2024})},\ \Eprint
  {http://arxiv.org/abs/2407.03501} {arXiv:2407.03501 [hep-ph]} \BibitemShut
  {NoStop}%
\bibitem [{\citenamefont {Ahrens}\ \emph {et~al.}(2010)\citenamefont {Ahrens},
  \citenamefont {Ferroglia}, \citenamefont {Neubert}, \citenamefont {Pecjak},\
  and\ \citenamefont {Yang}}]{Ahrens:2010zv}%
  \BibitemOpen
  \bibfield  {author} {\bibinfo {author} {\bibfnamefont {V.}~\bibnamefont
  {Ahrens}}, \bibinfo {author} {\bibfnamefont {A.}~\bibnamefont {Ferroglia}},
  \bibinfo {author} {\bibfnamefont {M.}~\bibnamefont {Neubert}}, \bibinfo
  {author} {\bibfnamefont {B.~D.}\ \bibnamefont {Pecjak}}, \ and\ \bibinfo
  {author} {\bibfnamefont {L.~L.}\ \bibnamefont {Yang}},\ }\href {\doibase
  10.1007/JHEP09(2010)097} {\bibfield  {journal} {\bibinfo  {journal} {JHEP}\
  }\textbf {\bibinfo {volume} {09}},\ \bibinfo {pages} {097} (\bibinfo {year}
  {2010})},\ \Eprint {http://arxiv.org/abs/1003.5827} {arXiv:1003.5827
  [hep-ph]} \BibitemShut {NoStop}%
\bibitem [{\citenamefont {Catani}\ \emph {et~al.}(2019)\citenamefont {Catani},
  \citenamefont {Devoto}, \citenamefont {Grazzini}, \citenamefont {Kallweit},
  \citenamefont {Mazzitelli},\ and\ \citenamefont {Sargsyan}}]{Catani:2019iny}%
  \BibitemOpen
  \bibfield  {author} {\bibinfo {author} {\bibfnamefont {S.}~\bibnamefont
  {Catani}}, \bibinfo {author} {\bibfnamefont {S.}~\bibnamefont {Devoto}},
  \bibinfo {author} {\bibfnamefont {M.}~\bibnamefont {Grazzini}}, \bibinfo
  {author} {\bibfnamefont {S.}~\bibnamefont {Kallweit}}, \bibinfo {author}
  {\bibfnamefont {J.}~\bibnamefont {Mazzitelli}}, \ and\ \bibinfo {author}
  {\bibfnamefont {H.}~\bibnamefont {Sargsyan}},\ }\href {\doibase
  10.1103/PhysRevD.99.051501} {\bibfield  {journal} {\bibinfo  {journal} {Phys.
  Rev. D}\ }\textbf {\bibinfo {volume} {99}},\ \bibinfo {pages} {051501}
  (\bibinfo {year} {2019})},\ \Eprint {http://arxiv.org/abs/1901.04005}
  {arXiv:1901.04005 [hep-ph]} \BibitemShut {NoStop}%
\bibitem [{\citenamefont {Catani}\ \emph
  {et~al.}(2023{\natexlab{a}})\citenamefont {Catani}, \citenamefont {Devoto},
  \citenamefont {Grazzini}, \citenamefont {Kallweit}, \citenamefont
  {Mazzitelli},\ and\ \citenamefont {Savoini}}]{Catani:2022mfv}%
  \BibitemOpen
  \bibfield  {author} {\bibinfo {author} {\bibfnamefont {S.}~\bibnamefont
  {Catani}}, \bibinfo {author} {\bibfnamefont {S.}~\bibnamefont {Devoto}},
  \bibinfo {author} {\bibfnamefont {M.}~\bibnamefont {Grazzini}}, \bibinfo
  {author} {\bibfnamefont {S.}~\bibnamefont {Kallweit}}, \bibinfo {author}
  {\bibfnamefont {J.}~\bibnamefont {Mazzitelli}}, \ and\ \bibinfo {author}
  {\bibfnamefont {C.}~\bibnamefont {Savoini}},\ }\href {\doibase
  10.1103/PhysRevLett.130.111902} {\bibfield  {journal} {\bibinfo  {journal}
  {Phys. Rev. Lett.}\ }\textbf {\bibinfo {volume} {130}},\ \bibinfo {pages}
  {111902} (\bibinfo {year} {2023}{\natexlab{a}})},\ \Eprint
  {http://arxiv.org/abs/2210.07846} {arXiv:2210.07846 [hep-ph]} \BibitemShut
  {NoStop}%
\bibitem [{\citenamefont {Buonocore}\ \emph {et~al.}(2023)\citenamefont
  {Buonocore}, \citenamefont {Devoto}, \citenamefont {Kallweit}, \citenamefont
  {Mazzitelli}, \citenamefont {Rottoli},\ and\ \citenamefont
  {Savoini}}]{Buonocore:2022pqq}%
  \BibitemOpen
  \bibfield  {author} {\bibinfo {author} {\bibfnamefont {L.}~\bibnamefont
  {Buonocore}}, \bibinfo {author} {\bibfnamefont {S.}~\bibnamefont {Devoto}},
  \bibinfo {author} {\bibfnamefont {S.}~\bibnamefont {Kallweit}}, \bibinfo
  {author} {\bibfnamefont {J.}~\bibnamefont {Mazzitelli}}, \bibinfo {author}
  {\bibfnamefont {L.}~\bibnamefont {Rottoli}}, \ and\ \bibinfo {author}
  {\bibfnamefont {C.}~\bibnamefont {Savoini}},\ }\href {\doibase
  10.1103/PhysRevD.107.074032} {\bibfield  {journal} {\bibinfo  {journal}
  {Phys. Rev. D}\ }\textbf {\bibinfo {volume} {107}},\ \bibinfo {pages}
  {074032} (\bibinfo {year} {2023})},\ \Eprint
  {http://arxiv.org/abs/2212.04954} {arXiv:2212.04954 [hep-ph]} \BibitemShut
  {NoStop}%
\bibitem [{\citenamefont {Devoto}\ \emph {et~al.}(2025)\citenamefont {Devoto},
  \citenamefont {Grazzini}, \citenamefont {Kallweit}, \citenamefont
  {Mazzitelli},\ and\ \citenamefont {Savoini}}]{Devoto:2024nhl}%
  \BibitemOpen
  \bibfield  {author} {\bibinfo {author} {\bibfnamefont {S.}~\bibnamefont
  {Devoto}}, \bibinfo {author} {\bibfnamefont {M.}~\bibnamefont {Grazzini}},
  \bibinfo {author} {\bibfnamefont {S.}~\bibnamefont {Kallweit}}, \bibinfo
  {author} {\bibfnamefont {J.}~\bibnamefont {Mazzitelli}}, \ and\ \bibinfo
  {author} {\bibfnamefont {C.}~\bibnamefont {Savoini}},\ }\href {\doibase
  10.1007/JHEP03(2025)189} {\bibfield  {journal} {\bibinfo  {journal} {JHEP}\
  }\textbf {\bibinfo {volume} {03}},\ \bibinfo {pages} {189} (\bibinfo {year}
  {2025})},\ \Eprint {http://arxiv.org/abs/2411.15340} {arXiv:2411.15340
  [hep-ph]} \BibitemShut {NoStop}%
\bibitem [{\citenamefont {Buonocore}\ \emph {et~al.}(2025)\citenamefont
  {Buonocore}, \citenamefont {Grazzini}, \citenamefont {Kallweit},
  \citenamefont {Lindert},\ and\ \citenamefont {Savoini}}]{Buonocore:2025fqs}%
  \BibitemOpen
  \bibfield  {author} {\bibinfo {author} {\bibfnamefont {L.}~\bibnamefont
  {Buonocore}}, \bibinfo {author} {\bibfnamefont {M.}~\bibnamefont {Grazzini}},
  \bibinfo {author} {\bibfnamefont {S.}~\bibnamefont {Kallweit}}, \bibinfo
  {author} {\bibfnamefont {J.~M.}\ \bibnamefont {Lindert}}, \ and\ \bibinfo
  {author} {\bibfnamefont {C.}~\bibnamefont {Savoini}},\ }\href {\doibase
  10.1007/JHEP10(2025)195} {\bibfield  {journal} {\bibinfo  {journal} {JHEP}\
  }\textbf {\bibinfo {volume} {10}},\ \bibinfo {pages} {195} (\bibinfo {year}
  {2025})},\ \Eprint {http://arxiv.org/abs/2507.11410} {arXiv:2507.11410
  [hep-ph]} \BibitemShut {NoStop}%
\bibitem [{\citenamefont {Mazzitelli}\ \emph {et~al.}(2021)\citenamefont
  {Mazzitelli}, \citenamefont {Monni}, \citenamefont {Nason}, \citenamefont
  {Re}, \citenamefont {Wiesemann},\ and\ \citenamefont
  {Zanderighi}}]{Mazzitelli:2020jio}%
  \BibitemOpen
  \bibfield  {author} {\bibinfo {author} {\bibfnamefont {J.}~\bibnamefont
  {Mazzitelli}}, \bibinfo {author} {\bibfnamefont {P.~F.}\ \bibnamefont
  {Monni}}, \bibinfo {author} {\bibfnamefont {P.}~\bibnamefont {Nason}},
  \bibinfo {author} {\bibfnamefont {E.}~\bibnamefont {Re}}, \bibinfo {author}
  {\bibfnamefont {M.}~\bibnamefont {Wiesemann}}, \ and\ \bibinfo {author}
  {\bibfnamefont {G.}~\bibnamefont {Zanderighi}},\ }\href {\doibase
  10.1103/PhysRevLett.127.062001} {\bibfield  {journal} {\bibinfo  {journal}
  {Phys. Rev. Lett.}\ }\textbf {\bibinfo {volume} {127}},\ \bibinfo {pages}
  {062001} (\bibinfo {year} {2021})},\ \Eprint
  {http://arxiv.org/abs/2012.14267} {arXiv:2012.14267 [hep-ph]} \BibitemShut
  {NoStop}%
\bibitem [{\citenamefont {Mazzitelli}\ \emph {et~al.}(2022)\citenamefont
  {Mazzitelli}, \citenamefont {Monni}, \citenamefont {Nason}, \citenamefont
  {Re}, \citenamefont {Wiesemann},\ and\ \citenamefont
  {Zanderighi}}]{Mazzitelli:2021mmm}%
  \BibitemOpen
  \bibfield  {author} {\bibinfo {author} {\bibfnamefont {J.}~\bibnamefont
  {Mazzitelli}}, \bibinfo {author} {\bibfnamefont {P.~F.}\ \bibnamefont
  {Monni}}, \bibinfo {author} {\bibfnamefont {P.}~\bibnamefont {Nason}},
  \bibinfo {author} {\bibfnamefont {E.}~\bibnamefont {Re}}, \bibinfo {author}
  {\bibfnamefont {M.}~\bibnamefont {Wiesemann}}, \ and\ \bibinfo {author}
  {\bibfnamefont {G.}~\bibnamefont {Zanderighi}},\ }\href {\doibase
  10.1007/JHEP04(2022)079} {\bibfield  {journal} {\bibinfo  {journal} {JHEP}\
  }\textbf {\bibinfo {volume} {04}},\ \bibinfo {pages} {079} (\bibinfo {year}
  {2022})},\ \Eprint {http://arxiv.org/abs/2112.12135} {arXiv:2112.12135
  [hep-ph]} \BibitemShut {NoStop}%
\bibitem [{\citenamefont {Mazzitelli}\ \emph {et~al.}(2024)\citenamefont
  {Mazzitelli}, \citenamefont {Sotnikov},\ and\ \citenamefont
  {Wiesemann}}]{Mazzitelli:2024ura}%
  \BibitemOpen
  \bibfield  {author} {\bibinfo {author} {\bibfnamefont {J.}~\bibnamefont
  {Mazzitelli}}, \bibinfo {author} {\bibfnamefont {V.}~\bibnamefont
  {Sotnikov}}, \ and\ \bibinfo {author} {\bibfnamefont {M.}~\bibnamefont
  {Wiesemann}},\ }\href@noop {} {\  (\bibinfo {year} {2024})},\ \Eprint
  {http://arxiv.org/abs/2404.08598} {arXiv:2404.08598 [hep-ph]} \BibitemShut
  {NoStop}%
\bibitem [{\citenamefont {Li}\ \emph {et~al.}(2020)\citenamefont {Li},
  \citenamefont {Neill},\ and\ \citenamefont {Zhu}}]{Li:2016axz}%
  \BibitemOpen
  \bibfield  {author} {\bibinfo {author} {\bibfnamefont {Y.}~\bibnamefont
  {Li}}, \bibinfo {author} {\bibfnamefont {D.}~\bibnamefont {Neill}}, \ and\
  \bibinfo {author} {\bibfnamefont {H.~X.}\ \bibnamefont {Zhu}},\ }\href
  {\doibase 10.1016/j.nuclphysb.2020.115193} {\bibfield  {journal} {\bibinfo
  {journal} {Nucl. Phys. B}\ }\textbf {\bibinfo {volume} {960}},\ \bibinfo
  {pages} {115193} (\bibinfo {year} {2020})},\ \Eprint
  {http://arxiv.org/abs/1604.00392} {arXiv:1604.00392 [hep-ph]} \BibitemShut
  {NoStop}%
\bibitem [{\citenamefont {Li}\ and\ \citenamefont {Zhu}(2017)}]{Li:2016ctv}%
  \BibitemOpen
  \bibfield  {author} {\bibinfo {author} {\bibfnamefont {Y.}~\bibnamefont
  {Li}}\ and\ \bibinfo {author} {\bibfnamefont {H.~X.}\ \bibnamefont {Zhu}},\
  }\href {\doibase 10.1103/PhysRevLett.118.022004} {\bibfield  {journal}
  {\bibinfo  {journal} {Phys. Rev. Lett.}\ }\textbf {\bibinfo {volume} {118}},\
  \bibinfo {pages} {022004} (\bibinfo {year} {2017})},\ \Eprint
  {http://arxiv.org/abs/1604.01404} {arXiv:1604.01404 [hep-ph]} \BibitemShut
  {NoStop}%
\bibitem [{\citenamefont {Liu}\ and\ \citenamefont
  {Monni}(2025{\natexlab{a}})}]{Liu:2024hfa}%
  \BibitemOpen
  \bibfield  {author} {\bibinfo {author} {\bibfnamefont {Z.~L.}\ \bibnamefont
  {Liu}}\ and\ \bibinfo {author} {\bibfnamefont {P.~F.}\ \bibnamefont
  {Monni}},\ }\href {\doibase 10.1007/JHEP03(2025)096} {\bibfield  {journal}
  {\bibinfo  {journal} {JHEP}\ }\textbf {\bibinfo {volume} {03}},\ \bibinfo
  {pages} {096} (\bibinfo {year} {2025}{\natexlab{a}})},\ \Eprint
  {http://arxiv.org/abs/2411.13466} {arXiv:2411.13466 [hep-ph]} \BibitemShut
  {NoStop}%
\bibitem [{\citenamefont {Czakon}\ and\ \citenamefont
  {Fiedler}(2014)}]{Czakon:2013hxa}%
  \BibitemOpen
  \bibfield  {author} {\bibinfo {author} {\bibfnamefont {M.}~\bibnamefont
  {Czakon}}\ and\ \bibinfo {author} {\bibfnamefont {P.}~\bibnamefont
  {Fiedler}},\ }\href {\doibase 10.1016/j.nuclphysb.2013.12.008} {\bibfield
  {journal} {\bibinfo  {journal} {Nucl. Phys. B}\ }\textbf {\bibinfo {volume}
  {879}},\ \bibinfo {pages} {236} (\bibinfo {year} {2014})},\ \Eprint
  {http://arxiv.org/abs/1311.2541} {arXiv:1311.2541 [hep-ph]} \BibitemShut
  {NoStop}%
\bibitem [{\citenamefont {Wang}\ \emph {et~al.}(2018)\citenamefont {Wang},
  \citenamefont {Xu}, \citenamefont {Yang},\ and\ \citenamefont
  {Zhu}}]{Wang:2018vgu}%
  \BibitemOpen
  \bibfield  {author} {\bibinfo {author} {\bibfnamefont {G.}~\bibnamefont
  {Wang}}, \bibinfo {author} {\bibfnamefont {X.}~\bibnamefont {Xu}}, \bibinfo
  {author} {\bibfnamefont {L.~L.}\ \bibnamefont {Yang}}, \ and\ \bibinfo
  {author} {\bibfnamefont {H.~X.}\ \bibnamefont {Zhu}},\ }\href {\doibase
  10.1007/JHEP06(2018)013} {\bibfield  {journal} {\bibinfo  {journal} {JHEP}\
  }\textbf {\bibinfo {volume} {06}},\ \bibinfo {pages} {013} (\bibinfo {year}
  {2018})},\ \Eprint {http://arxiv.org/abs/1804.05218} {arXiv:1804.05218
  [hep-ph]} \BibitemShut {NoStop}%
\bibitem [{\citenamefont {Angeles-Martinez}\ \emph {et~al.}(2018)\citenamefont
  {Angeles-Martinez}, \citenamefont {Czakon},\ and\ \citenamefont
  {Sapeta}}]{Angeles-Martinez:2018mqh}%
  \BibitemOpen
  \bibfield  {author} {\bibinfo {author} {\bibfnamefont {R.}~\bibnamefont
  {Angeles-Martinez}}, \bibinfo {author} {\bibfnamefont {M.}~\bibnamefont
  {Czakon}}, \ and\ \bibinfo {author} {\bibfnamefont {S.}~\bibnamefont
  {Sapeta}},\ }\href {\doibase 10.1007/JHEP10(2018)201} {\bibfield  {journal}
  {\bibinfo  {journal} {JHEP}\ }\textbf {\bibinfo {volume} {10}},\ \bibinfo
  {pages} {201} (\bibinfo {year} {2018})},\ \Eprint
  {http://arxiv.org/abs/1809.01459} {arXiv:1809.01459 [hep-ph]} \BibitemShut
  {NoStop}%
\bibitem [{\citenamefont {Catani}\ \emph
  {et~al.}(2023{\natexlab{b}})\citenamefont {Catani}, \citenamefont {Devoto},
  \citenamefont {Grazzini},\ and\ \citenamefont {Mazzitelli}}]{Catani:2023tby}%
  \BibitemOpen
  \bibfield  {author} {\bibinfo {author} {\bibfnamefont {S.}~\bibnamefont
  {Catani}}, \bibinfo {author} {\bibfnamefont {S.}~\bibnamefont {Devoto}},
  \bibinfo {author} {\bibfnamefont {M.}~\bibnamefont {Grazzini}}, \ and\
  \bibinfo {author} {\bibfnamefont {J.}~\bibnamefont {Mazzitelli}},\ }\href
  {\doibase 10.1007/JHEP04(2023)144} {\bibfield  {journal} {\bibinfo  {journal}
  {JHEP}\ }\textbf {\bibinfo {volume} {04}},\ \bibinfo {pages} {144} (\bibinfo
  {year} {2023}{\natexlab{b}})},\ \Eprint {http://arxiv.org/abs/2301.11786}
  {arXiv:2301.11786 [hep-ph]} \BibitemShut {NoStop}%
\bibitem [{\citenamefont {Devoto}\ and\ \citenamefont
  {Mazzitelli}(2025)}]{Devoto:2025eyc}%
  \BibitemOpen
  \bibfield  {author} {\bibinfo {author} {\bibfnamefont {S.}~\bibnamefont
  {Devoto}}\ and\ \bibinfo {author} {\bibfnamefont {J.}~\bibnamefont
  {Mazzitelli}},\ }\href@noop {} {\  (\bibinfo {year} {2025})},\ \Eprint
  {http://arxiv.org/abs/2509.17509} {arXiv:2509.17509 [hep-ph]} \BibitemShut
  {NoStop}%
\bibitem [{\citenamefont {Shao}\ and\ \citenamefont
  {Wang}(2025{\natexlab{a}})}]{Shao:2025dzw}%
  \BibitemOpen
  \bibfield  {author} {\bibinfo {author} {\bibfnamefont {H.-S.}\ \bibnamefont
  {Shao}}\ and\ \bibinfo {author} {\bibfnamefont {G.}~\bibnamefont {Wang}},\
  }\href {\doibase 10.1007/JHEP10(2025)164} {\bibfield  {journal} {\bibinfo
  {journal} {JHEP}\ }\textbf {\bibinfo {volume} {10}},\ \bibinfo {pages} {164}
  (\bibinfo {year} {2025}{\natexlab{a}})},\ \Eprint
  {http://arxiv.org/abs/2506.23791} {arXiv:2506.23791 [hep-ph]} \BibitemShut
  {NoStop}%
\bibitem [{\citenamefont {Catani}\ \emph {et~al.}(2017)\citenamefont {Catani},
  \citenamefont {Grazzini},\ and\ \citenamefont {Sargsyan}}]{Catani:2017tuc}%
  \BibitemOpen
  \bibfield  {author} {\bibinfo {author} {\bibfnamefont {S.}~\bibnamefont
  {Catani}}, \bibinfo {author} {\bibfnamefont {M.}~\bibnamefont {Grazzini}}, \
  and\ \bibinfo {author} {\bibfnamefont {H.}~\bibnamefont {Sargsyan}},\ }\href
  {\doibase 10.1007/JHEP06(2017)017} {\bibfield  {journal} {\bibinfo  {journal}
  {JHEP}\ }\textbf {\bibinfo {volume} {06}},\ \bibinfo {pages} {017} (\bibinfo
  {year} {2017})},\ \Eprint {http://arxiv.org/abs/1703.08468} {arXiv:1703.08468
  [hep-ph]} \BibitemShut {NoStop}%
\bibitem [{\citenamefont {Liu}\ and\ \citenamefont
  {Monni}(2025{\natexlab{b}})}]{zenodo}%
  \BibitemOpen
  \bibfield  {author} {\bibinfo {author} {\bibfnamefont {Z.~L.}\ \bibnamefont
  {Liu}}\ and\ \bibinfo {author} {\bibfnamefont {P.~F.}\ \bibnamefont
  {Monni}},\ }\href {\doibase 10.5281/zenodo.17739950} {\enquote {\bibinfo
  {title} {Ancillary files provided with the submission of this article},}\ }
  (\bibinfo {year} {2025}{\natexlab{b}})\BibitemShut {NoStop}%
\bibitem [{\citenamefont {Collins}\ \emph {et~al.}(1985)\citenamefont
  {Collins}, \citenamefont {Soper},\ and\ \citenamefont
  {Sterman}}]{Collins:1984kg}%
  \BibitemOpen
  \bibfield  {author} {\bibinfo {author} {\bibfnamefont {J.~C.}\ \bibnamefont
  {Collins}}, \bibinfo {author} {\bibfnamefont {D.~E.}\ \bibnamefont {Soper}},
  \ and\ \bibinfo {author} {\bibfnamefont {G.~F.}\ \bibnamefont {Sterman}},\
  }\href {\doibase 10.1016/0550-3213(85)90479-1} {\bibfield  {journal}
  {\bibinfo  {journal} {Nucl. Phys. B}\ }\textbf {\bibinfo {volume} {250}},\
  \bibinfo {pages} {199} (\bibinfo {year} {1985})}\BibitemShut {NoStop}%
\bibitem [{\citenamefont {Ji}\ \emph {et~al.}(2005)\citenamefont {Ji},
  \citenamefont {Ma},\ and\ \citenamefont {Yuan}}]{Ji:2004wu}%
  \BibitemOpen
  \bibfield  {author} {\bibinfo {author} {\bibfnamefont {X.-d.}\ \bibnamefont
  {Ji}}, \bibinfo {author} {\bibfnamefont {J.-p.}\ \bibnamefont {Ma}}, \ and\
  \bibinfo {author} {\bibfnamefont {F.}~\bibnamefont {Yuan}},\ }\href {\doibase
  10.1103/PhysRevD.71.034005} {\bibfield  {journal} {\bibinfo  {journal} {Phys.
  Rev. D}\ }\textbf {\bibinfo {volume} {71}},\ \bibinfo {pages} {034005}
  (\bibinfo {year} {2005})},\ \Eprint {http://arxiv.org/abs/hep-ph/0404183}
  {arXiv:hep-ph/0404183} \BibitemShut {NoStop}%
\bibitem [{\citenamefont {Becher}\ and\ \citenamefont
  {Neubert}(2011)}]{Becher:2010tm}%
  \BibitemOpen
  \bibfield  {author} {\bibinfo {author} {\bibfnamefont {T.}~\bibnamefont
  {Becher}}\ and\ \bibinfo {author} {\bibfnamefont {M.}~\bibnamefont
  {Neubert}},\ }\href {\doibase 10.1140/epjc/s10052-011-1665-7} {\bibfield
  {journal} {\bibinfo  {journal} {Eur. Phys. J. C}\ }\textbf {\bibinfo {volume}
  {71}},\ \bibinfo {pages} {1665} (\bibinfo {year} {2011})},\ \Eprint
  {http://arxiv.org/abs/1007.4005} {arXiv:1007.4005 [hep-ph]} \BibitemShut
  {NoStop}%
\bibitem [{\citenamefont {Chiu}\ \emph {et~al.}(2012)\citenamefont {Chiu},
  \citenamefont {Jain}, \citenamefont {Neill},\ and\ \citenamefont
  {Rothstein}}]{Chiu:2011qc}%
  \BibitemOpen
  \bibfield  {author} {\bibinfo {author} {\bibfnamefont {J.-y.}\ \bibnamefont
  {Chiu}}, \bibinfo {author} {\bibfnamefont {A.}~\bibnamefont {Jain}}, \bibinfo
  {author} {\bibfnamefont {D.}~\bibnamefont {Neill}}, \ and\ \bibinfo {author}
  {\bibfnamefont {I.~Z.}\ \bibnamefont {Rothstein}},\ }\href {\doibase
  10.1103/PhysRevLett.108.151601} {\bibfield  {journal} {\bibinfo  {journal}
  {Phys. Rev. Lett.}\ }\textbf {\bibinfo {volume} {108}},\ \bibinfo {pages}
  {151601} (\bibinfo {year} {2012})},\ \Eprint {http://arxiv.org/abs/1104.0881}
  {arXiv:1104.0881 [hep-ph]} \BibitemShut {NoStop}%
\bibitem [{\citenamefont {Bauer}\ \emph
  {et~al.}(2002{\natexlab{a}})\citenamefont {Bauer}, \citenamefont {Pirjol},\
  and\ \citenamefont {Stewart}}]{Bauer:2001yt}%
  \BibitemOpen
  \bibfield  {author} {\bibinfo {author} {\bibfnamefont {C.~W.}\ \bibnamefont
  {Bauer}}, \bibinfo {author} {\bibfnamefont {D.}~\bibnamefont {Pirjol}}, \
  and\ \bibinfo {author} {\bibfnamefont {I.~W.}\ \bibnamefont {Stewart}},\
  }\href {\doibase 10.1103/PhysRevD.65.054022} {\bibfield  {journal} {\bibinfo
  {journal} {Phys. Rev. D}\ }\textbf {\bibinfo {volume} {65}},\ \bibinfo
  {pages} {054022} (\bibinfo {year} {2002}{\natexlab{a}})},\ \Eprint
  {http://arxiv.org/abs/hep-ph/0109045} {arXiv:hep-ph/0109045} \BibitemShut
  {NoStop}%
\bibitem [{\citenamefont {Bauer}\ \emph
  {et~al.}(2002{\natexlab{b}})\citenamefont {Bauer}, \citenamefont {Fleming},
  \citenamefont {Pirjol}, \citenamefont {Rothstein},\ and\ \citenamefont
  {Stewart}}]{Bauer:2002nz}%
  \BibitemOpen
  \bibfield  {author} {\bibinfo {author} {\bibfnamefont {C.~W.}\ \bibnamefont
  {Bauer}}, \bibinfo {author} {\bibfnamefont {S.}~\bibnamefont {Fleming}},
  \bibinfo {author} {\bibfnamefont {D.}~\bibnamefont {Pirjol}}, \bibinfo
  {author} {\bibfnamefont {I.~Z.}\ \bibnamefont {Rothstein}}, \ and\ \bibinfo
  {author} {\bibfnamefont {I.~W.}\ \bibnamefont {Stewart}},\ }\href {\doibase
  10.1103/PhysRevD.66.014017} {\bibfield  {journal} {\bibinfo  {journal} {Phys.
  Rev. D}\ }\textbf {\bibinfo {volume} {66}},\ \bibinfo {pages} {014017}
  (\bibinfo {year} {2002}{\natexlab{b}})},\ \Eprint
  {http://arxiv.org/abs/hep-ph/0202088} {arXiv:hep-ph/0202088} \BibitemShut
  {NoStop}%
\bibitem [{\citenamefont {Beneke}\ \emph {et~al.}(2002)\citenamefont {Beneke},
  \citenamefont {Chapovsky}, \citenamefont {Diehl},\ and\ \citenamefont
  {Feldmann}}]{Beneke:2002ph}%
  \BibitemOpen
  \bibfield  {author} {\bibinfo {author} {\bibfnamefont {M.}~\bibnamefont
  {Beneke}}, \bibinfo {author} {\bibfnamefont {A.~P.}\ \bibnamefont
  {Chapovsky}}, \bibinfo {author} {\bibfnamefont {M.}~\bibnamefont {Diehl}}, \
  and\ \bibinfo {author} {\bibfnamefont {T.}~\bibnamefont {Feldmann}},\ }\href
  {\doibase 10.1016/S0550-3213(02)00687-9} {\bibfield  {journal} {\bibinfo
  {journal} {Nucl. Phys. B}\ }\textbf {\bibinfo {volume} {643}},\ \bibinfo
  {pages} {431} (\bibinfo {year} {2002})},\ \Eprint
  {http://arxiv.org/abs/hep-ph/0206152} {arXiv:hep-ph/0206152} \BibitemShut
  {NoStop}%
\bibitem [{\citenamefont {Luo}\ \emph {et~al.}(2019)\citenamefont {Luo},
  \citenamefont {Wang}, \citenamefont {Xu}, \citenamefont {Yang}, \citenamefont
  {Yang},\ and\ \citenamefont {Zhu}}]{Luo:2019hmp}%
  \BibitemOpen
  \bibfield  {author} {\bibinfo {author} {\bibfnamefont {M.-X.}\ \bibnamefont
  {Luo}}, \bibinfo {author} {\bibfnamefont {X.}~\bibnamefont {Wang}}, \bibinfo
  {author} {\bibfnamefont {X.}~\bibnamefont {Xu}}, \bibinfo {author}
  {\bibfnamefont {L.~L.}\ \bibnamefont {Yang}}, \bibinfo {author}
  {\bibfnamefont {T.-Z.}\ \bibnamefont {Yang}}, \ and\ \bibinfo {author}
  {\bibfnamefont {H.~X.}\ \bibnamefont {Zhu}},\ }\href {\doibase
  10.1007/JHEP10(2019)083} {\bibfield  {journal} {\bibinfo  {journal} {JHEP}\
  }\textbf {\bibinfo {volume} {10}},\ \bibinfo {pages} {083} (\bibinfo {year}
  {2019})},\ \Eprint {http://arxiv.org/abs/1908.03831} {arXiv:1908.03831
  [hep-ph]} \BibitemShut {NoStop}%
\bibitem [{\citenamefont {Luo}\ \emph {et~al.}(2020{\natexlab{a}})\citenamefont
  {Luo}, \citenamefont {Yang}, \citenamefont {Zhu},\ and\ \citenamefont
  {Zhu}}]{Luo:2019bmw}%
  \BibitemOpen
  \bibfield  {author} {\bibinfo {author} {\bibfnamefont {M.-X.}\ \bibnamefont
  {Luo}}, \bibinfo {author} {\bibfnamefont {T.-Z.}\ \bibnamefont {Yang}},
  \bibinfo {author} {\bibfnamefont {H.~X.}\ \bibnamefont {Zhu}}, \ and\
  \bibinfo {author} {\bibfnamefont {Y.~J.}\ \bibnamefont {Zhu}},\ }\href
  {\doibase 10.1007/JHEP01(2020)040} {\bibfield  {journal} {\bibinfo  {journal}
  {JHEP}\ }\textbf {\bibinfo {volume} {01}},\ \bibinfo {pages} {040} (\bibinfo
  {year} {2020}{\natexlab{a}})},\ \Eprint {http://arxiv.org/abs/1909.13820}
  {arXiv:1909.13820 [hep-ph]} \BibitemShut {NoStop}%
\bibitem [{\citenamefont {Luo}\ \emph {et~al.}(2020{\natexlab{b}})\citenamefont
  {Luo}, \citenamefont {Yang}, \citenamefont {Zhu},\ and\ \citenamefont
  {Zhu}}]{Luo:2019szz}%
  \BibitemOpen
  \bibfield  {author} {\bibinfo {author} {\bibfnamefont {M.-x.}\ \bibnamefont
  {Luo}}, \bibinfo {author} {\bibfnamefont {T.-Z.}\ \bibnamefont {Yang}},
  \bibinfo {author} {\bibfnamefont {H.~X.}\ \bibnamefont {Zhu}}, \ and\
  \bibinfo {author} {\bibfnamefont {Y.~J.}\ \bibnamefont {Zhu}},\ }\href
  {\doibase 10.1103/PhysRevLett.124.092001} {\bibfield  {journal} {\bibinfo
  {journal} {Phys. Rev. Lett.}\ }\textbf {\bibinfo {volume} {124}},\ \bibinfo
  {pages} {092001} (\bibinfo {year} {2020}{\natexlab{b}})},\ \Eprint
  {http://arxiv.org/abs/1912.05778} {arXiv:1912.05778 [hep-ph]} \BibitemShut
  {NoStop}%
\bibitem [{\citenamefont {Ebert}\ \emph {et~al.}(2020)\citenamefont {Ebert},
  \citenamefont {Mistlberger},\ and\ \citenamefont {Vita}}]{Ebert:2020yqt}%
  \BibitemOpen
  \bibfield  {author} {\bibinfo {author} {\bibfnamefont {M.~A.}\ \bibnamefont
  {Ebert}}, \bibinfo {author} {\bibfnamefont {B.}~\bibnamefont {Mistlberger}},
  \ and\ \bibinfo {author} {\bibfnamefont {G.}~\bibnamefont {Vita}},\ }\href
  {\doibase 10.1007/JHEP09(2020)146} {\bibfield  {journal} {\bibinfo  {journal}
  {JHEP}\ }\textbf {\bibinfo {volume} {09}},\ \bibinfo {pages} {146} (\bibinfo
  {year} {2020})},\ \Eprint {http://arxiv.org/abs/2006.05329} {arXiv:2006.05329
  [hep-ph]} \BibitemShut {NoStop}%
\bibitem [{\citenamefont {Luo}\ \emph {et~al.}(2021)\citenamefont {Luo},
  \citenamefont {Yang}, \citenamefont {Zhu},\ and\ \citenamefont
  {Zhu}}]{Luo:2020epw}%
  \BibitemOpen
  \bibfield  {author} {\bibinfo {author} {\bibfnamefont {M.-x.}\ \bibnamefont
  {Luo}}, \bibinfo {author} {\bibfnamefont {T.-Z.}\ \bibnamefont {Yang}},
  \bibinfo {author} {\bibfnamefont {H.~X.}\ \bibnamefont {Zhu}}, \ and\
  \bibinfo {author} {\bibfnamefont {Y.~J.}\ \bibnamefont {Zhu}},\ }\href
  {\doibase 10.1007/JHEP06(2021)115} {\bibfield  {journal} {\bibinfo  {journal}
  {JHEP}\ }\textbf {\bibinfo {volume} {06}},\ \bibinfo {pages} {115} (\bibinfo
  {year} {2021})},\ \Eprint {http://arxiv.org/abs/2012.03256} {arXiv:2012.03256
  [hep-ph]} \BibitemShut {NoStop}%
\bibitem [{\citenamefont {Shao}\ and\ \citenamefont
  {Wang}(2025{\natexlab{b}})}]{Shao:2025qgv}%
  \BibitemOpen
  \bibfield  {author} {\bibinfo {author} {\bibfnamefont {H.-S.}\ \bibnamefont
  {Shao}}\ and\ \bibinfo {author} {\bibfnamefont {G.}~\bibnamefont {Wang}},\
  }\href {\doibase 10.1007/JHEP06(2025)255} {\bibfield  {journal} {\bibinfo
  {journal} {JHEP}\ }\textbf {\bibinfo {volume} {06}},\ \bibinfo {pages} {255}
  (\bibinfo {year} {2025}{\natexlab{b}})},\ \Eprint
  {http://arxiv.org/abs/2504.05131} {arXiv:2504.05131 [hep-ph]} \BibitemShut
  {NoStop}%
\bibitem [{\citenamefont {Liu}\ and\ \citenamefont
  {Schalch}(2022)}]{Liu:2022elt}%
  \BibitemOpen
  \bibfield  {author} {\bibinfo {author} {\bibfnamefont {Z.~L.}\ \bibnamefont
  {Liu}}\ and\ \bibinfo {author} {\bibfnamefont {N.}~\bibnamefont {Schalch}},\
  }\href {\doibase 10.1103/PhysRevLett.129.232001} {\bibfield  {journal}
  {\bibinfo  {journal} {Phys. Rev. Lett.}\ }\textbf {\bibinfo {volume} {129}},\
  \bibinfo {pages} {232001} (\bibinfo {year} {2022})},\ \Eprint
  {http://arxiv.org/abs/2207.02864} {arXiv:2207.02864 [hep-ph]} \BibitemShut
  {NoStop}%
\bibitem [{\citenamefont {Gardi}\ and\ \citenamefont
  {Zhu}(2025)}]{Gardi:2025lws}%
  \BibitemOpen
  \bibfield  {author} {\bibinfo {author} {\bibfnamefont {E.}~\bibnamefont
  {Gardi}}\ and\ \bibinfo {author} {\bibfnamefont {Z.}~\bibnamefont {Zhu}},\
  }\href@noop {} {\  (\bibinfo {year} {2025})},\ \Eprint
  {http://arxiv.org/abs/2510.27567} {arXiv:2510.27567 [hep-ph]} \BibitemShut
  {NoStop}%
\bibitem [{\citenamefont {Collins}(2003)}]{Collins:2003fm}%
  \BibitemOpen
  \bibfield  {author} {\bibinfo {author} {\bibfnamefont {J.~C.}\ \bibnamefont
  {Collins}},\ }\href@noop {} {\bibfield  {journal} {\bibinfo  {journal} {Acta
  Phys. Polon. B}\ }\textbf {\bibinfo {volume} {34}},\ \bibinfo {pages} {3103}
  (\bibinfo {year} {2003})},\ \Eprint {http://arxiv.org/abs/hep-ph/0304122}
  {arXiv:hep-ph/0304122} \BibitemShut {NoStop}%
\bibitem [{\citenamefont {Becher}\ and\ \citenamefont
  {Bell}(2012)}]{Becher:2011dz}%
  \BibitemOpen
  \bibfield  {author} {\bibinfo {author} {\bibfnamefont {T.}~\bibnamefont
  {Becher}}\ and\ \bibinfo {author} {\bibfnamefont {G.}~\bibnamefont {Bell}},\
  }\href {\doibase 10.1016/j.physletb.2012.05.016} {\bibfield  {journal}
  {\bibinfo  {journal} {Phys. Lett. B}\ }\textbf {\bibinfo {volume} {713}},\
  \bibinfo {pages} {41} (\bibinfo {year} {2012})},\ \Eprint
  {http://arxiv.org/abs/1112.3907} {arXiv:1112.3907 [hep-ph]} \BibitemShut
  {NoStop}%
\bibitem [{\citenamefont {Buonocore}\ \emph {et~al.}(2022)\citenamefont
  {Buonocore}, \citenamefont {Grazzini}, \citenamefont {Haag}, \citenamefont
  {Rottoli},\ and\ \citenamefont {Savoini}}]{Buonocore:2022mle}%
  \BibitemOpen
  \bibfield  {author} {\bibinfo {author} {\bibfnamefont {L.}~\bibnamefont
  {Buonocore}}, \bibinfo {author} {\bibfnamefont {M.}~\bibnamefont {Grazzini}},
  \bibinfo {author} {\bibfnamefont {J.}~\bibnamefont {Haag}}, \bibinfo {author}
  {\bibfnamefont {L.}~\bibnamefont {Rottoli}}, \ and\ \bibinfo {author}
  {\bibfnamefont {C.}~\bibnamefont {Savoini}},\ }\href {\doibase
  10.1103/PhysRevD.106.014008} {\bibfield  {journal} {\bibinfo  {journal}
  {Phys. Rev. D}\ }\textbf {\bibinfo {volume} {106}},\ \bibinfo {pages}
  {014008} (\bibinfo {year} {2022})},\ \Eprint
  {http://arxiv.org/abs/2201.11519} {arXiv:2201.11519 [hep-ph]} \BibitemShut
  {NoStop}%
\bibitem [{\citenamefont {Abreu}\ \emph {et~al.}(2022)\citenamefont {Abreu},
  \citenamefont {Gaunt}, \citenamefont {Monni},\ and\ \citenamefont
  {Szafron}}]{Abreu:2022sdc}%
  \BibitemOpen
  \bibfield  {author} {\bibinfo {author} {\bibfnamefont {S.}~\bibnamefont
  {Abreu}}, \bibinfo {author} {\bibfnamefont {J.~R.}\ \bibnamefont {Gaunt}},
  \bibinfo {author} {\bibfnamefont {P.~F.}\ \bibnamefont {Monni}}, \ and\
  \bibinfo {author} {\bibfnamefont {R.}~\bibnamefont {Szafron}},\ }\href
  {\doibase 10.1007/JHEP08(2022)268} {\bibfield  {journal} {\bibinfo  {journal}
  {JHEP}\ }\textbf {\bibinfo {volume} {08}},\ \bibinfo {pages} {268} (\bibinfo
  {year} {2022})},\ \Eprint {http://arxiv.org/abs/2204.02987} {arXiv:2204.02987
  [hep-ph]} \BibitemShut {NoStop}%
\bibitem [{\citenamefont {Fu}\ \emph {et~al.}(2025)\citenamefont {Fu},
  \citenamefont {Rahn}, \citenamefont {Shao}, \citenamefont {Waalewijn},\ and\
  \citenamefont {Wu}}]{Fu:2024fgj}%
  \BibitemOpen
  \bibfield  {author} {\bibinfo {author} {\bibfnamefont {R.-J.}\ \bibnamefont
  {Fu}}, \bibinfo {author} {\bibfnamefont {R.}~\bibnamefont {Rahn}}, \bibinfo
  {author} {\bibfnamefont {D.~Y.}\ \bibnamefont {Shao}}, \bibinfo {author}
  {\bibfnamefont {W.~J.}\ \bibnamefont {Waalewijn}}, \ and\ \bibinfo {author}
  {\bibfnamefont {B.}~\bibnamefont {Wu}},\ }\href {\doibase 10.1103/htvz-wz1p}
  {\bibfield  {journal} {\bibinfo  {journal} {Phys. Rev. Lett.}\ }\textbf
  {\bibinfo {volume} {135}},\ \bibinfo {pages} {171903} (\bibinfo {year}
  {2025})},\ \Eprint {http://arxiv.org/abs/2412.05358} {arXiv:2412.05358
  [hep-ph]} \BibitemShut {NoStop}%
\bibitem [{\citenamefont {van Beekveld}\ \emph {et~al.}(2025)\citenamefont {van
  Beekveld}, \citenamefont {Buonocore}, \citenamefont {Ferrario~Ravasio},
  \citenamefont {Monni}, \citenamefont {Soto-Ontoso},\ and\ \citenamefont
  {Soyez}}]{vanBeekveld:2025zjh}%
  \BibitemOpen
  \bibfield  {author} {\bibinfo {author} {\bibfnamefont {M.}~\bibnamefont {van
  Beekveld}}, \bibinfo {author} {\bibfnamefont {L.}~\bibnamefont {Buonocore}},
  \bibinfo {author} {\bibfnamefont {S.}~\bibnamefont {Ferrario~Ravasio}},
  \bibinfo {author} {\bibfnamefont {P.~F.}\ \bibnamefont {Monni}}, \bibinfo
  {author} {\bibfnamefont {A.}~\bibnamefont {Soto-Ontoso}}, \ and\ \bibinfo
  {author} {\bibfnamefont {G.}~\bibnamefont {Soyez}},\ }\href@noop {} {\
  (\bibinfo {year} {2025})},\ \Eprint {http://arxiv.org/abs/2511.16723}
  {arXiv:2511.16723 [hep-ph]} \BibitemShut {NoStop}%
\bibitem [{\citenamefont {Li}\ \emph {et~al.}(2011)\citenamefont {Li},
  \citenamefont {Mantry},\ and\ \citenamefont {Petriello}}]{Li:2011zp}%
  \BibitemOpen
  \bibfield  {author} {\bibinfo {author} {\bibfnamefont {Y.}~\bibnamefont
  {Li}}, \bibinfo {author} {\bibfnamefont {S.}~\bibnamefont {Mantry}}, \ and\
  \bibinfo {author} {\bibfnamefont {F.}~\bibnamefont {Petriello}},\ }\href
  {\doibase 10.1103/PhysRevD.84.094014} {\bibfield  {journal} {\bibinfo
  {journal} {Phys. Rev. D}\ }\textbf {\bibinfo {volume} {84}},\ \bibinfo
  {pages} {094014} (\bibinfo {year} {2011})},\ \Eprint
  {http://arxiv.org/abs/1105.5171} {arXiv:1105.5171 [hep-ph]} \BibitemShut
  {NoStop}%
\bibitem [{\citenamefont {Catani}\ and\ \citenamefont
  {Seymour}(1996)}]{Catani:1996jh}%
  \BibitemOpen
  \bibfield  {author} {\bibinfo {author} {\bibfnamefont {S.}~\bibnamefont
  {Catani}}\ and\ \bibinfo {author} {\bibfnamefont {M.~H.}\ \bibnamefont
  {Seymour}},\ }\href {\doibase 10.1016/0370-2693(96)00425-X} {\bibfield
  {journal} {\bibinfo  {journal} {Phys. Lett. B}\ }\textbf {\bibinfo {volume}
  {378}},\ \bibinfo {pages} {287} (\bibinfo {year} {1996})},\ \Eprint
  {http://arxiv.org/abs/hep-ph/9602277} {arXiv:hep-ph/9602277} \BibitemShut
  {NoStop}%
\bibitem [{\citenamefont {Catani}\ and\ \citenamefont
  {Seymour}(1997)}]{Catani:1996vz}%
  \BibitemOpen
  \bibfield  {author} {\bibinfo {author} {\bibfnamefont {S.}~\bibnamefont
  {Catani}}\ and\ \bibinfo {author} {\bibfnamefont {M.~H.}\ \bibnamefont
  {Seymour}},\ }\href {\doibase 10.1016/S0550-3213(96)00589-5} {\bibfield
  {journal} {\bibinfo  {journal} {Nucl. Phys. B}\ }\textbf {\bibinfo {volume}
  {485}},\ \bibinfo {pages} {291} (\bibinfo {year} {1997})},\ \bibinfo {note}
  {[Erratum: Nucl.Phys.B 510, 503--504 (1998)]},\ \Eprint
  {http://arxiv.org/abs/hep-ph/9605323} {arXiv:hep-ph/9605323} \BibitemShut
  {NoStop}%
\bibitem [{\citenamefont {Catani}\ and\ \citenamefont
  {Grazzini}(2000{\natexlab{a}})}]{Catani:1999ss}%
  \BibitemOpen
  \bibfield  {author} {\bibinfo {author} {\bibfnamefont {S.}~\bibnamefont
  {Catani}}\ and\ \bibinfo {author} {\bibfnamefont {M.}~\bibnamefont
  {Grazzini}},\ }\href {\doibase 10.1016/S0550-3213(99)00778-6} {\bibfield
  {journal} {\bibinfo  {journal} {Nucl. Phys. B}\ }\textbf {\bibinfo {volume}
  {570}},\ \bibinfo {pages} {287} (\bibinfo {year} {2000}{\natexlab{a}})},\
  \Eprint {http://arxiv.org/abs/hep-ph/9908523} {arXiv:hep-ph/9908523}
  \BibitemShut {NoStop}%
\bibitem [{\citenamefont {Catani}\ and\ \citenamefont
  {Grazzini}(2000{\natexlab{b}})}]{Catani:2000pi}%
  \BibitemOpen
  \bibfield  {author} {\bibinfo {author} {\bibfnamefont {S.}~\bibnamefont
  {Catani}}\ and\ \bibinfo {author} {\bibfnamefont {M.}~\bibnamefont
  {Grazzini}},\ }\href {\doibase 10.1016/S0550-3213(00)00572-1} {\bibfield
  {journal} {\bibinfo  {journal} {Nucl. Phys. B}\ }\textbf {\bibinfo {volume}
  {591}},\ \bibinfo {pages} {435} (\bibinfo {year} {2000}{\natexlab{b}})},\
  \Eprint {http://arxiv.org/abs/hep-ph/0007142} {arXiv:hep-ph/0007142}
  \BibitemShut {NoStop}%
\bibitem [{\citenamefont {Czakon}(2011)}]{Czakon:2011ve}%
  \BibitemOpen
  \bibfield  {author} {\bibinfo {author} {\bibfnamefont {M.}~\bibnamefont
  {Czakon}},\ }\href {\doibase 10.1016/j.nuclphysb.2011.03.020} {\bibfield
  {journal} {\bibinfo  {journal} {Nucl. Phys. B}\ }\textbf {\bibinfo {volume}
  {849}},\ \bibinfo {pages} {250} (\bibinfo {year} {2011})},\ \Eprint
  {http://arxiv.org/abs/1101.0642} {arXiv:1101.0642 [hep-ph]} \BibitemShut
  {NoStop}%
\bibitem [{\citenamefont {Bierenbaum}\ \emph {et~al.}(2012)\citenamefont
  {Bierenbaum}, \citenamefont {Czakon},\ and\ \citenamefont
  {Mitov}}]{Bierenbaum:2011gg}%
  \BibitemOpen
  \bibfield  {author} {\bibinfo {author} {\bibfnamefont {I.}~\bibnamefont
  {Bierenbaum}}, \bibinfo {author} {\bibfnamefont {M.}~\bibnamefont {Czakon}},
  \ and\ \bibinfo {author} {\bibfnamefont {A.}~\bibnamefont {Mitov}},\ }\href
  {\doibase 10.1016/j.nuclphysb.2011.11.002} {\bibfield  {journal} {\bibinfo
  {journal} {Nucl. Phys. B}\ }\textbf {\bibinfo {volume} {856}},\ \bibinfo
  {pages} {228} (\bibinfo {year} {2012})},\ \Eprint
  {http://arxiv.org/abs/1107.4384} {arXiv:1107.4384 [hep-ph]} \BibitemShut
  {NoStop}%
\bibitem [{\citenamefont {Czakon}\ and\ \citenamefont
  {Mitov}(2018)}]{Czakon:2018iev}%
  \BibitemOpen
  \bibfield  {author} {\bibinfo {author} {\bibfnamefont {M.~L.}\ \bibnamefont
  {Czakon}}\ and\ \bibinfo {author} {\bibfnamefont {A.}~\bibnamefont {Mitov}},\
  }\href@noop {} {\  (\bibinfo {year} {2018})},\ \Eprint
  {http://arxiv.org/abs/1804.02069} {arXiv:1804.02069 [hep-ph]} \BibitemShut
  {NoStop}%
\bibitem [{\citenamefont {Anastasiou}\ and\ \citenamefont
  {Melnikov}(2002)}]{Anastasiou:2002yz}%
  \BibitemOpen
  \bibfield  {author} {\bibinfo {author} {\bibfnamefont {C.}~\bibnamefont
  {Anastasiou}}\ and\ \bibinfo {author} {\bibfnamefont {K.}~\bibnamefont
  {Melnikov}},\ }\href {\doibase 10.1016/S0550-3213(02)00837-4} {\bibfield
  {journal} {\bibinfo  {journal} {Nucl. Phys. B}\ }\textbf {\bibinfo {volume}
  {646}},\ \bibinfo {pages} {220} (\bibinfo {year} {2002})},\ \Eprint
  {http://arxiv.org/abs/hep-ph/0207004} {arXiv:hep-ph/0207004} \BibitemShut
  {NoStop}%
\bibitem [{\citenamefont {Anastasiou}\ \emph {et~al.}(2003)\citenamefont
  {Anastasiou}, \citenamefont {Dixon}, \citenamefont {Melnikov},\ and\
  \citenamefont {Petriello}}]{Anastasiou:2003yy}%
  \BibitemOpen
  \bibfield  {author} {\bibinfo {author} {\bibfnamefont {C.}~\bibnamefont
  {Anastasiou}}, \bibinfo {author} {\bibfnamefont {L.~J.}\ \bibnamefont
  {Dixon}}, \bibinfo {author} {\bibfnamefont {K.}~\bibnamefont {Melnikov}}, \
  and\ \bibinfo {author} {\bibfnamefont {F.}~\bibnamefont {Petriello}},\ }\href
  {\doibase 10.1103/PhysRevLett.91.182002} {\bibfield  {journal} {\bibinfo
  {journal} {Phys. Rev. Lett.}\ }\textbf {\bibinfo {volume} {91}},\ \bibinfo
  {pages} {182002} (\bibinfo {year} {2003})},\ \Eprint
  {http://arxiv.org/abs/hep-ph/0306192} {arXiv:hep-ph/0306192} \BibitemShut
  {NoStop}%
\bibitem [{\citenamefont {Anastasiou}\ \emph {et~al.}(2004)\citenamefont
  {Anastasiou}, \citenamefont {Dixon}, \citenamefont {Melnikov},\ and\
  \citenamefont {Petriello}}]{Anastasiou:2003ds}%
  \BibitemOpen
  \bibfield  {author} {\bibinfo {author} {\bibfnamefont {C.}~\bibnamefont
  {Anastasiou}}, \bibinfo {author} {\bibfnamefont {L.~J.}\ \bibnamefont
  {Dixon}}, \bibinfo {author} {\bibfnamefont {K.}~\bibnamefont {Melnikov}}, \
  and\ \bibinfo {author} {\bibfnamefont {F.}~\bibnamefont {Petriello}},\ }\href
  {\doibase 10.1103/PhysRevD.69.094008} {\bibfield  {journal} {\bibinfo
  {journal} {Phys. Rev. D}\ }\textbf {\bibinfo {volume} {69}},\ \bibinfo
  {pages} {094008} (\bibinfo {year} {2004})},\ \Eprint
  {http://arxiv.org/abs/hep-ph/0312266} {arXiv:hep-ph/0312266} \BibitemShut
  {NoStop}%
\bibitem [{\citenamefont {Smirnov}\ and\ \citenamefont
  {Chukharev}(2020)}]{Smirnov:2019qkx}%
  \BibitemOpen
  \bibfield  {author} {\bibinfo {author} {\bibfnamefont {A.~V.}\ \bibnamefont
  {Smirnov}}\ and\ \bibinfo {author} {\bibfnamefont {F.~S.}\ \bibnamefont
  {Chukharev}},\ }\href {\doibase 10.1016/j.cpc.2019.106877} {\bibfield
  {journal} {\bibinfo  {journal} {Comput. Phys. Commun.}\ }\textbf {\bibinfo
  {volume} {247}},\ \bibinfo {pages} {106877} (\bibinfo {year} {2020})},\
  \Eprint {http://arxiv.org/abs/1901.07808} {arXiv:1901.07808 [hep-ph]}
  \BibitemShut {NoStop}%
\bibitem [{\citenamefont {Klappert}\ \emph {et~al.}(2021)\citenamefont
  {Klappert}, \citenamefont {Lange}, \citenamefont {Maierh\"ofer},\ and\
  \citenamefont {Usovitsch}}]{Klappert:2020nbg}%
  \BibitemOpen
  \bibfield  {author} {\bibinfo {author} {\bibfnamefont {J.}~\bibnamefont
  {Klappert}}, \bibinfo {author} {\bibfnamefont {F.}~\bibnamefont {Lange}},
  \bibinfo {author} {\bibfnamefont {P.}~\bibnamefont {Maierh\"ofer}}, \ and\
  \bibinfo {author} {\bibfnamefont {J.}~\bibnamefont {Usovitsch}},\ }\href
  {\doibase 10.1016/j.cpc.2021.108024} {\bibfield  {journal} {\bibinfo
  {journal} {Comput. Phys. Commun.}\ }\textbf {\bibinfo {volume} {266}},\
  \bibinfo {pages} {108024} (\bibinfo {year} {2021})},\ \Eprint
  {http://arxiv.org/abs/2008.06494} {arXiv:2008.06494 [hep-ph]} \BibitemShut
  {NoStop}%
\bibitem [{\citenamefont {Kotikov}(1991{\natexlab{a}})}]{Kotikov:1990kg}%
  \BibitemOpen
  \bibfield  {author} {\bibinfo {author} {\bibfnamefont {A.~V.}\ \bibnamefont
  {Kotikov}},\ }\href {\doibase 10.1016/0370-2693(91)90413-K} {\bibfield
  {journal} {\bibinfo  {journal} {Phys. Lett. B}\ }\textbf {\bibinfo {volume}
  {254}},\ \bibinfo {pages} {158} (\bibinfo {year}
  {1991}{\natexlab{a}})}\BibitemShut {NoStop}%
\bibitem [{\citenamefont {Kotikov}(1991{\natexlab{b}})}]{Kotikov:1991hm}%
  \BibitemOpen
  \bibfield  {author} {\bibinfo {author} {\bibfnamefont {A.~V.}\ \bibnamefont
  {Kotikov}},\ }\href {\doibase 10.1016/0370-2693(91)90834-D} {\bibfield
  {journal} {\bibinfo  {journal} {Phys. Lett. B}\ }\textbf {\bibinfo {volume}
  {259}},\ \bibinfo {pages} {314} (\bibinfo {year}
  {1991}{\natexlab{b}})}\BibitemShut {NoStop}%
\bibitem [{\citenamefont {Kotikov}(1991{\natexlab{c}})}]{Kotikov:1991pm}%
  \BibitemOpen
  \bibfield  {author} {\bibinfo {author} {\bibfnamefont {A.~V.}\ \bibnamefont
  {Kotikov}},\ }\href {\doibase 10.1016/0370-2693(91)90536-Y} {\bibfield
  {journal} {\bibinfo  {journal} {Phys. Lett. B}\ }\textbf {\bibinfo {volume}
  {267}},\ \bibinfo {pages} {123} (\bibinfo {year} {1991}{\natexlab{c}})},\
  \bibinfo {note} {[Erratum: Phys.Lett.B 295, 409--409 (1992)]}\BibitemShut
  {NoStop}%
\bibitem [{\citenamefont {Bern}\ \emph {et~al.}(1993)\citenamefont {Bern},
  \citenamefont {Dixon},\ and\ \citenamefont {Kosower}}]{Bern:1992em}%
  \BibitemOpen
  \bibfield  {author} {\bibinfo {author} {\bibfnamefont {Z.}~\bibnamefont
  {Bern}}, \bibinfo {author} {\bibfnamefont {L.~J.}\ \bibnamefont {Dixon}}, \
  and\ \bibinfo {author} {\bibfnamefont {D.~A.}\ \bibnamefont {Kosower}},\
  }\href {\doibase 10.1016/0370-2693(93)90400-C} {\bibfield  {journal}
  {\bibinfo  {journal} {Phys. Lett. B}\ }\textbf {\bibinfo {volume} {302}},\
  \bibinfo {pages} {299} (\bibinfo {year} {1993})},\ \bibinfo {note} {[Erratum:
  Phys.Lett.B 318, 649 (1993)]},\ \Eprint {http://arxiv.org/abs/hep-ph/9212308}
  {arXiv:hep-ph/9212308} \BibitemShut {NoStop}%
\bibitem [{\citenamefont {Bern}\ \emph {et~al.}(1994)\citenamefont {Bern},
  \citenamefont {Dixon},\ and\ \citenamefont {Kosower}}]{Bern:1993kr}%
  \BibitemOpen
  \bibfield  {author} {\bibinfo {author} {\bibfnamefont {Z.}~\bibnamefont
  {Bern}}, \bibinfo {author} {\bibfnamefont {L.~J.}\ \bibnamefont {Dixon}}, \
  and\ \bibinfo {author} {\bibfnamefont {D.~A.}\ \bibnamefont {Kosower}},\
  }\href {\doibase 10.1016/0550-3213(94)90398-0} {\bibfield  {journal}
  {\bibinfo  {journal} {Nucl. Phys. B}\ }\textbf {\bibinfo {volume} {412}},\
  \bibinfo {pages} {751} (\bibinfo {year} {1994})},\ \Eprint
  {http://arxiv.org/abs/hep-ph/9306240} {arXiv:hep-ph/9306240} \BibitemShut
  {NoStop}%
\bibitem [{\citenamefont {Remiddi}(1997)}]{Remiddi:1997ny}%
  \BibitemOpen
  \bibfield  {author} {\bibinfo {author} {\bibfnamefont {E.}~\bibnamefont
  {Remiddi}},\ }\href {\doibase 10.1007/BF03185566} {\bibfield  {journal}
  {\bibinfo  {journal} {Nuovo Cim. A}\ }\textbf {\bibinfo {volume} {110}},\
  \bibinfo {pages} {1435} (\bibinfo {year} {1997})},\ \Eprint
  {http://arxiv.org/abs/hep-th/9711188} {arXiv:hep-th/9711188} \BibitemShut
  {NoStop}%
\bibitem [{\citenamefont {Gehrmann}\ and\ \citenamefont
  {Remiddi}(2000)}]{Gehrmann:1999as}%
  \BibitemOpen
  \bibfield  {author} {\bibinfo {author} {\bibfnamefont {T.}~\bibnamefont
  {Gehrmann}}\ and\ \bibinfo {author} {\bibfnamefont {E.}~\bibnamefont
  {Remiddi}},\ }\href {\doibase 10.1016/S0550-3213(00)00223-6} {\bibfield
  {journal} {\bibinfo  {journal} {Nucl. Phys. B}\ }\textbf {\bibinfo {volume}
  {580}},\ \bibinfo {pages} {485} (\bibinfo {year} {2000})},\ \Eprint
  {http://arxiv.org/abs/hep-ph/9912329} {arXiv:hep-ph/9912329} \BibitemShut
  {NoStop}%
\bibitem [{\citenamefont {Lee}(2014)}]{Lee:2013mka}%
  \BibitemOpen
  \bibfield  {author} {\bibinfo {author} {\bibfnamefont {R.~N.}\ \bibnamefont
  {Lee}},\ }\href {\doibase 10.1088/1742-6596/523/1/012059} {\bibfield
  {journal} {\bibinfo  {journal} {J. Phys. Conf. Ser.}\ }\textbf {\bibinfo
  {volume} {523}},\ \bibinfo {pages} {012059} (\bibinfo {year} {2014})},\
  \Eprint {http://arxiv.org/abs/1310.1145} {arXiv:1310.1145 [hep-ph]}
  \BibitemShut {NoStop}%
\bibitem [{\citenamefont {Henn}(2013)}]{Henn:2013pwa}%
  \BibitemOpen
  \bibfield  {author} {\bibinfo {author} {\bibfnamefont {J.~M.}\ \bibnamefont
  {Henn}},\ }\href {\doibase 10.1103/PhysRevLett.110.251601} {\bibfield
  {journal} {\bibinfo  {journal} {Phys. Rev. Lett.}\ }\textbf {\bibinfo
  {volume} {110}},\ \bibinfo {pages} {251601} (\bibinfo {year} {2013})},\
  \Eprint {http://arxiv.org/abs/1304.1806} {arXiv:1304.1806 [hep-th]}
  \BibitemShut {NoStop}%
\bibitem [{\citenamefont {Argeri}\ \emph {et~al.}(2014)\citenamefont {Argeri},
  \citenamefont {Di~Vita}, \citenamefont {Mastrolia}, \citenamefont
  {Mirabella}, \citenamefont {Schlenk}, \citenamefont {Schubert},\ and\
  \citenamefont {Tancredi}}]{Argeri:2014qva}%
  \BibitemOpen
  \bibfield  {author} {\bibinfo {author} {\bibfnamefont {M.}~\bibnamefont
  {Argeri}}, \bibinfo {author} {\bibfnamefont {S.}~\bibnamefont {Di~Vita}},
  \bibinfo {author} {\bibfnamefont {P.}~\bibnamefont {Mastrolia}}, \bibinfo
  {author} {\bibfnamefont {E.}~\bibnamefont {Mirabella}}, \bibinfo {author}
  {\bibfnamefont {J.}~\bibnamefont {Schlenk}}, \bibinfo {author} {\bibfnamefont
  {U.}~\bibnamefont {Schubert}}, \ and\ \bibinfo {author} {\bibfnamefont
  {L.}~\bibnamefont {Tancredi}},\ }\href {\doibase 10.1007/JHEP03(2014)082}
  {\bibfield  {journal} {\bibinfo  {journal} {JHEP}\ }\textbf {\bibinfo
  {volume} {03}},\ \bibinfo {pages} {082} (\bibinfo {year} {2014})},\ \Eprint
  {http://arxiv.org/abs/1401.2979} {arXiv:1401.2979 [hep-ph]} \BibitemShut
  {NoStop}%
\bibitem [{\citenamefont {Cachazo}(2008)}]{Cachazo:2008vp}%
  \BibitemOpen
  \bibfield  {author} {\bibinfo {author} {\bibfnamefont {F.}~\bibnamefont
  {Cachazo}},\ }\href@noop {} {\  (\bibinfo {year} {2008})},\ \Eprint
  {http://arxiv.org/abs/0803.1988} {arXiv:0803.1988 [hep-th]} \BibitemShut
  {NoStop}%
\bibitem [{\citenamefont {Arkani-Hamed}\ \emph {et~al.}(2012)\citenamefont
  {Arkani-Hamed}, \citenamefont {Bourjaily}, \citenamefont {Cachazo},\ and\
  \citenamefont {Trnka}}]{Arkani-Hamed:2010pyv}%
  \BibitemOpen
  \bibfield  {author} {\bibinfo {author} {\bibfnamefont {N.}~\bibnamefont
  {Arkani-Hamed}}, \bibinfo {author} {\bibfnamefont {J.~L.}\ \bibnamefont
  {Bourjaily}}, \bibinfo {author} {\bibfnamefont {F.}~\bibnamefont {Cachazo}},
  \ and\ \bibinfo {author} {\bibfnamefont {J.}~\bibnamefont {Trnka}},\ }\href
  {\doibase 10.1007/JHEP06(2012)125} {\bibfield  {journal} {\bibinfo  {journal}
  {JHEP}\ }\textbf {\bibinfo {volume} {06}},\ \bibinfo {pages} {125} (\bibinfo
  {year} {2012})},\ \Eprint {http://arxiv.org/abs/1012.6032} {arXiv:1012.6032
  [hep-th]} \BibitemShut {NoStop}%
\bibitem [{\citenamefont {Liu}\ \emph {et~al.}(2018)\citenamefont {Liu},
  \citenamefont {Ma},\ and\ \citenamefont {Wang}}]{Liu:2017jxz}%
  \BibitemOpen
  \bibfield  {author} {\bibinfo {author} {\bibfnamefont {X.}~\bibnamefont
  {Liu}}, \bibinfo {author} {\bibfnamefont {Y.-Q.}\ \bibnamefont {Ma}}, \ and\
  \bibinfo {author} {\bibfnamefont {C.-Y.}\ \bibnamefont {Wang}},\ }\href
  {\doibase 10.1016/j.physletb.2018.02.026} {\bibfield  {journal} {\bibinfo
  {journal} {Phys. Lett. B}\ }\textbf {\bibinfo {volume} {779}},\ \bibinfo
  {pages} {353} (\bibinfo {year} {2018})},\ \Eprint
  {http://arxiv.org/abs/1711.09572} {arXiv:1711.09572 [hep-ph]} \BibitemShut
  {NoStop}%
\bibitem [{\citenamefont {Liu}\ and\ \citenamefont {Ma}(2023)}]{Liu:2022chg}%
  \BibitemOpen
  \bibfield  {author} {\bibinfo {author} {\bibfnamefont {X.}~\bibnamefont
  {Liu}}\ and\ \bibinfo {author} {\bibfnamefont {Y.-Q.}\ \bibnamefont {Ma}},\
  }\href {\doibase 10.1016/j.cpc.2022.108565} {\bibfield  {journal} {\bibinfo
  {journal} {Comput. Phys. Commun.}\ }\textbf {\bibinfo {volume} {283}},\
  \bibinfo {pages} {108565} (\bibinfo {year} {2023})},\ \Eprint
  {http://arxiv.org/abs/2201.11669} {arXiv:2201.11669 [hep-ph]} \BibitemShut
  {NoStop}%
\bibitem [{\citenamefont {Besier}\ \emph {et~al.}(2020)\citenamefont {Besier},
  \citenamefont {Wasser},\ and\ \citenamefont {Weinzierl}}]{Besier:2019kco}%
  \BibitemOpen
  \bibfield  {author} {\bibinfo {author} {\bibfnamefont {M.}~\bibnamefont
  {Besier}}, \bibinfo {author} {\bibfnamefont {P.}~\bibnamefont {Wasser}}, \
  and\ \bibinfo {author} {\bibfnamefont {S.}~\bibnamefont {Weinzierl}},\ }\href
  {\doibase 10.1016/j.cpc.2020.107197} {\bibfield  {journal} {\bibinfo
  {journal} {Comput. Phys. Commun.}\ }\textbf {\bibinfo {volume} {253}},\
  \bibinfo {pages} {107197} (\bibinfo {year} {2020})},\ \Eprint
  {http://arxiv.org/abs/1910.13251} {arXiv:1910.13251 [cs.MS]} \BibitemShut
  {NoStop}%
\bibitem [{\citenamefont {Becher}\ and\ \citenamefont
  {Neubert}(2009{\natexlab{a}})}]{Becher:2009cu}%
  \BibitemOpen
  \bibfield  {author} {\bibinfo {author} {\bibfnamefont {T.}~\bibnamefont
  {Becher}}\ and\ \bibinfo {author} {\bibfnamefont {M.}~\bibnamefont
  {Neubert}},\ }\href {\doibase 10.1103/PhysRevLett.102.162001,
  10.1103/PhysRevLett.111.199905} {\bibfield  {journal} {\bibinfo  {journal}
  {Phys. Rev. Lett.}\ }\textbf {\bibinfo {volume} {102}},\ \bibinfo {pages}
  {162001} (\bibinfo {year} {2009}{\natexlab{a}})},\ \bibinfo {note} {[Erratum:
  Phys. Rev. Lett.111,no.19,199905(2013)]},\ \Eprint
  {http://arxiv.org/abs/0901.0722} {arXiv:0901.0722 [hep-ph]} \BibitemShut
  {NoStop}%
%%CITATION = ARXIV:0901.0722;%%
\bibitem [{\citenamefont {Becher}\ and\ \citenamefont
  {Neubert}(2009{\natexlab{b}})}]{Becher:2009qa}%
  \BibitemOpen
  \bibfield  {author} {\bibinfo {author} {\bibfnamefont {T.}~\bibnamefont
  {Becher}}\ and\ \bibinfo {author} {\bibfnamefont {M.}~\bibnamefont
  {Neubert}},\ }\href {\doibase 10.1088/1126-6708/2009/06/081,
  10.1007/JHEP11(2013)024} {\bibfield  {journal} {\bibinfo  {journal} {JHEP}\
  }\textbf {\bibinfo {volume} {06}},\ \bibinfo {pages} {081} (\bibinfo {year}
  {2009}{\natexlab{b}})},\ \bibinfo {note} {[Erratum: JHEP11,024(2013)]},\
  \Eprint {http://arxiv.org/abs/0903.1126} {arXiv:0903.1126 [hep-ph]}
  \BibitemShut {NoStop}%
%%CITATION = ARXIV:0903.1126;%%
\bibitem [{\citenamefont {Becher}\ and\ \citenamefont
  {Neubert}(2009{\natexlab{c}})}]{Becher:2009kw}%
  \BibitemOpen
  \bibfield  {author} {\bibinfo {author} {\bibfnamefont {T.}~\bibnamefont
  {Becher}}\ and\ \bibinfo {author} {\bibfnamefont {M.}~\bibnamefont
  {Neubert}},\ }\href {\doibase 10.1103/PhysRevD.79.125004} {\bibfield
  {journal} {\bibinfo  {journal} {Phys. Rev. D}\ }\textbf {\bibinfo {volume}
  {79}},\ \bibinfo {pages} {125004} (\bibinfo {year} {2009}{\natexlab{c}})},\
  \bibinfo {note} {[Erratum: Phys.Rev.D 80, 109901(E) (2009)]},\ \Eprint
  {http://arxiv.org/abs/0904.1021} {arXiv:0904.1021 [hep-ph]} \BibitemShut
  {NoStop}%
\bibitem [{\citenamefont {Ferroglia}\ \emph
  {et~al.}(2009{\natexlab{a}})\citenamefont {Ferroglia}, \citenamefont
  {Neubert}, \citenamefont {Pecjak},\ and\ \citenamefont
  {Yang}}]{Ferroglia:2009ep}%
  \BibitemOpen
  \bibfield  {author} {\bibinfo {author} {\bibfnamefont {A.}~\bibnamefont
  {Ferroglia}}, \bibinfo {author} {\bibfnamefont {M.}~\bibnamefont {Neubert}},
  \bibinfo {author} {\bibfnamefont {B.~D.}\ \bibnamefont {Pecjak}}, \ and\
  \bibinfo {author} {\bibfnamefont {L.~L.}\ \bibnamefont {Yang}},\ }\href
  {\doibase 10.1103/PhysRevLett.103.201601} {\bibfield  {journal} {\bibinfo
  {journal} {Phys. Rev. Lett.}\ }\textbf {\bibinfo {volume} {103}},\ \bibinfo
  {pages} {201601} (\bibinfo {year} {2009}{\natexlab{a}})},\ \Eprint
  {http://arxiv.org/abs/0907.4791} {arXiv:0907.4791 [hep-ph]} \BibitemShut
  {NoStop}%
\bibitem [{\citenamefont {Kidonakis}(2009)}]{Kidonakis:2009ev}%
  \BibitemOpen
  \bibfield  {author} {\bibinfo {author} {\bibfnamefont {N.}~\bibnamefont
  {Kidonakis}},\ }\href {\doibase 10.1103/PhysRevLett.102.232003} {\bibfield
  {journal} {\bibinfo  {journal} {Phys. Rev. Lett.}\ }\textbf {\bibinfo
  {volume} {102}},\ \bibinfo {pages} {232003} (\bibinfo {year} {2009})},\
  \Eprint {http://arxiv.org/abs/0903.2561} {arXiv:0903.2561 [hep-ph]}
  \BibitemShut {NoStop}%
\bibitem [{\citenamefont {Grozin}\ \emph {et~al.}(2015)\citenamefont {Grozin},
  \citenamefont {Henn}, \citenamefont {Korchemsky},\ and\ \citenamefont
  {Marquard}}]{Grozin:2014hna}%
  \BibitemOpen
  \bibfield  {author} {\bibinfo {author} {\bibfnamefont {A.}~\bibnamefont
  {Grozin}}, \bibinfo {author} {\bibfnamefont {J.~M.}\ \bibnamefont {Henn}},
  \bibinfo {author} {\bibfnamefont {G.~P.}\ \bibnamefont {Korchemsky}}, \ and\
  \bibinfo {author} {\bibfnamefont {P.}~\bibnamefont {Marquard}},\ }\href
  {\doibase 10.1103/PhysRevLett.114.062006} {\bibfield  {journal} {\bibinfo
  {journal} {Phys. Rev. Lett.}\ }\textbf {\bibinfo {volume} {114}},\ \bibinfo
  {pages} {062006} (\bibinfo {year} {2015})},\ \Eprint
  {http://arxiv.org/abs/1409.0023} {arXiv:1409.0023 [hep-ph]} \BibitemShut
  {NoStop}%
\bibitem [{\citenamefont {Ferroglia}\ \emph
  {et~al.}(2009{\natexlab{b}})\citenamefont {Ferroglia}, \citenamefont
  {Neubert}, \citenamefont {Pecjak},\ and\ \citenamefont
  {Yang}}]{Ferroglia:2009ii}%
  \BibitemOpen
  \bibfield  {author} {\bibinfo {author} {\bibfnamefont {A.}~\bibnamefont
  {Ferroglia}}, \bibinfo {author} {\bibfnamefont {M.}~\bibnamefont {Neubert}},
  \bibinfo {author} {\bibfnamefont {B.~D.}\ \bibnamefont {Pecjak}}, \ and\
  \bibinfo {author} {\bibfnamefont {L.~L.}\ \bibnamefont {Yang}},\ }\href
  {\doibase 10.1088/1126-6708/2009/11/062} {\bibfield  {journal} {\bibinfo
  {journal} {JHEP}\ }\textbf {\bibinfo {volume} {11}},\ \bibinfo {pages} {062}
  (\bibinfo {year} {2009}{\natexlab{b}})},\ \Eprint
  {http://arxiv.org/abs/0908.3676} {arXiv:0908.3676 [hep-ph]} \BibitemShut
  {NoStop}%
\end{thebibliography}%

%========================
%=== Supplemental material ===
%========================

\onecolumngrid
\newpage
\appendix

\section*{Supplemental material}

\subsection{A. Rationalization for the differential equations}\label{app:scfac}
The square roots encountered in the DEs for the tripole contribution to ${\boldsymbol \cT}_{12I}$ are  
%\begin{widetext}
\begin{equation}\label{eq:srootsA}
R^A_1=\sqrt{\kappa_I^2-1}\,,\qquad 
R^A_2=\sqrt{\frac{\left(t_{1I}-t_{2I}\right)^2}{t_{1I}^2 t_{2I}^2}- \frac{4(\kappa_I \left(t_{1I}+t_{2I}\right)-t_{1I} t_{2I}-1)}{\chi \,t_{1I}t_{2I}}
+\frac{4 \left(\kappa_I^2-1\right)}{\chi^2} }\,,
\end{equation}
%\end{widetext}
which can be rationalized by the following variable changes
\begin{gather}
\kappa_I=\frac{(1-a_I)^2+1}{2(1-a_I)}\,,\qquad
\chi=\frac{ x^2 \left(t_{1I}-t_{2I}\right)^2+4 t_{1I}^2 t_{2I}^2 -4 t_{1I}^2 t_{2I}^2 \kappa_I^2 }{2\left[x\left(t_{1I}-t_{2I}\right)^2
+2 t_{1I} t_{2I} \left(t_{1I}t_{2I}+1\right)
-2 t_{1I} t_{2I} \kappa_I \left(t_{1I}+t_{2I}\right)\right]}\,.
\end{gather}
For the tripole contribution to ${\boldsymbol \cT}_{j34}$, there are four square roots involved  
\begin{gather}
R^B_1=\sqrt{\rho^2-1}\,,\qquad
R^B_2=\sqrt{\kappa_3^2-1}\,,\qquad
R^B_3=\sqrt{\kappa_4^2-1}\,,\nn\\
R^B_4=\sqrt{\frac{2 \left(\rho -\kappa_3 \kappa _4\right)}{t_{j3} t_{j4}}
+\frac{2 \left(\kappa_3-\kappa_4 \rho\right)}{t_{j3}}
+\frac{2 \left(\kappa_4-\kappa_3 \rho\right)}{t_{j4}}
+\frac{\kappa_3^2-1}{t_{j4}^2}+\frac{\kappa_4^2-1}{t_{j3}^2}
+\rho ^2-1}\,,\label{eq:srootsB}
\end{gather}
which can be rationalized by the following variable changes
\begin{gather}
\rho=\frac{(1-r)^2+1}{2(1-r)}\,,\qquad
\kappa_I=\frac{(1-a_I)^2+1}{2(1-a_I)}\,,\qquad
t_{j3}=\frac{z_{j}}{u_{34}}\,,\qquad 
t_{j4}=\frac{1}{u_{34} z_{j}}\,,\nn\\
u_{34}=\frac{2\left[(\kappa _4-\kappa _3 \rho) z_j+(\kappa _3-\kappa _4 \rho )z_j^{-1}+( \rho ^2-1) y_{34}\right]}
{(\rho ^2-1)y_{34}^2 + 2 (\kappa _3 \kappa _4-\rho) -\left(\kappa _3^2-1\right) z_j^2 - (\kappa _4^2-1)z_j^{-2} }\,.
\end{gather}

%\renewcommand{\theequation}{A\arabic{equation}}
%\setcounter{equation}{0}
%============ appendix B========
%\subsection{B. Dlog form for the differential equations}\label{app:m0andcollim}

\end{document}